\documentclass[11pt]{article}
\usepackage{axodraw}
\usepackage{epsfig}
\newcommand{\la}[1]{\label{#1}}
\newcommand{\be}{\begin{equation}}
\newcommand{\ee}{\end{equation}}
\newcommand{\ba}{\begin{eqnarray}}
\newcommand{\ea}{\end{eqnarray}}
\newcommand{\rmi}[1]{{\mbox{\scriptsize #1}}}

\newcommand{\eq}{Eq.~}
\newcommand{\se}{Sec.~}
\newcommand{\eqs}{Eqs.~}
\newcommand{\nr}[1]{(\ref{#1})}
\newcommand{\tr}{{\rm Tr\,}}
\newcommand{\nn}{\nonumber \\}
\newcommand{\fr}[2]{{\frac{#1}{#2}\,}}

\renewcommand{\vec}[1]{{\bf #1}}

\newcommand{\Ms}{M_{\rm s}}

\newcommand{\Nc}{N_{\rm c}}

\newcommand{\rmO}{{\mathcal{O}}}

\def\lsi{\raise0.3ex\hbox{$<$\kern-0.75em\raise-1.1ex\hbox{$\sim$}}}
\def\gsi{\raise0.3ex\hbox{$>$\kern-0.75em\raise-1.1ex\hbox{$\sim$}}}
\newcommand{\lsim}{\mathop{\lsi}}
\newcommand{\gsim}{\mathop{\gsi}}

\newcommand{\fe}{\rmi{f}}
\newcommand{\bo}{\rmi{b}}
\newcommand{\disc}{\mathop{\mbox{Disc}}}
\newcommand{\nF}[1]{n_\rmi{F{#1}}}
\newcommand{\nB}{n_\rmi{B}}
\newcommand{\rmii}[1]{{\mbox{\tiny\rm{#1}}}}
\newcommand{\re}{\mathop{\mbox{Re}}}
\newcommand{\im}{\mathop{\mbox{Im}}}
\newcommand{\bsl}[1]{\,\slash\!\!\!\!{#1}\,}
\newcommand{\msl}[1]{\,\slash\!\!\!{#1}\,}
\newcommand{\Tint}[1]{{\hbox{$\sum$}\!\!\!\!\!\!\int}_{\!\!\!\!\raise-0.9ex\hbox{$\scriptstyle{#1}$}}}

\newcommand{\pic}[1]{\;\parbox[c]{30pt}{\begin{picture}(30,30)(0,0)
\SetWidth{1.0}\SetScale{1.0} #1 \end{picture}}\;}

\newcommand{\picc}[1]{\;\parbox[c]{60pt}{\begin{picture}(60,30)(0,0)
\SetWidth{1.0}\SetScale{1.0} #1 \end{picture}}\;}

\newcommand{\piccc}[1]{\;\parbox[c]{90pt}{\begin{picture}(90,30)(0,0)
\SetWidth{1.0}\SetScale{1.0} #1 \end{picture}}\;}

\def\Lwidth{1}

\def\Agl(#1,#2)(#3,#4,#5){\PhotonArc(#1,#2)(#3,#4,#5){\Lwidth}
{6.283 #3 mul 360 div #4 #5 sub #4 #5 sub mul sqrt mul Ldensity mul}}
\def\Lgl(#1,#2)(#3,#4){\Photon(#1,#2)(#3,#4){\Lwidth}
{#1 #3 sub #1 #3 sub mul #2 #4 sub #2 #4 sub mul add sqrt Ldensity mul}}
\def\Agh(#1,#2)(#3,#4,#5){\DashArrowArc(#1,#2)(#3,#4,#5){1}}
\def\Aagh(#1,#2)(#3,#4,#5){\DashArrowArcn(#1,#2)(#3,#5,#4){1}}
\def\Lgh(#1,#2)(#3,#4){\DashArrowLine(#1,#2)(#3,#4){1}}
\def\Lagh(#1,#2)(#3,#4){\DashArrowLine(#3,#4)(#1,#2){1}}
\def\Ahh(#1,#2)(#3,#4,#5){\DashCArc(#1,#2)(#3,#4,#5){1}}
\def\Lhh(#1,#2)(#3,#4){\DashLine(#1,#2)(#3,#4){1}}
\def\Aqu(#1,#2)(#3,#4,#5){\ArrowArc(#1,#2)(#3,#4,#5)}
\def\Aaqu(#1,#2)(#3,#4,#5){\ArrowArcn(#1,#2)(#3,#5,#4)}
\def\Lqu(#1,#2)(#3,#4){\ArrowLine(#1,#2)(#3,#4)}
\def\Laqu(#1,#2)(#3,#4){\ArrowLine(#3,#4)(#1,#2)}
\def\Aqq(#1,#2)(#3,#4,#5){\CArc(#1,#2)(#3,#4,#5)}
\def\Lqq(#1,#2)(#3,#4){\Line(#1,#2)(#3,#4)}
\def\Asc(#1,#2)(#3,#4,#5){\CArc(#1,#2)(#3,#4,#5)}
\def\Lsc(#1,#2)(#3,#4){\Line(#1,#2)(#3,#4)}

\def\scfc{0.7}  
\def\phgt{21}   
\def\pwc{21}    
\def\pwcc{42} 
\newcommand{\PIC}[4]{\;\parbox[c]{#2 pt}{\begin{picture}(#2,#3)(0,0)
\SetWidth{1.0}\SetScale{#4} #1 \end{picture}}\;}
\renewcommand{\pic}[1]{\PIC{#1}{\pwc}{\phgt}{\scfc}}

\renewcommand{\picc}[1]{\PIC{#1}{\pwcc}{\phgt}{\scfc}}
\def\Textt(#1,#2,#3){\Text(#1,#2)[t]{{$\scriptstyle #3$}}}
\def\Textb(#1,#2,#3){\Text(#1,#2)[b]{{$\scriptstyle #3$}}}
\def\Textl(#1,#2,#3){\Text(#1,#2)[l]{{$\scriptstyle #3$}}}
\def\Textr(#1,#2,#3){\Text(#1,#2)[r]{{$\scriptstyle #3$}}}
\def\Texttl(#1,#2,#3){\Text(#1,#2)[tl]{{$\scriptstyle #3$}}}
\def\Textbl(#1,#2,#3){\Text(#1,#2)[bl]{{$\scriptstyle #3$}}}
\def\Texttr(#1,#2,#3){\Text(#1,#2)[tr]{{$\scriptstyle #3$}}}
\def\Textbr(#1,#2,#3){\Text(#1,#2)[br]{{$\scriptstyle #3$}}}
\def\Textst(#1,#2,#3){\Text(#1,#2)[t]{{$\scriptscriptstyle #3$}}}
\def\Textsb(#1,#2,#3){\Text(#1,#2)[b]{{$\scriptscriptstyle #3$}}}
\def\Textsl(#1,#2,#3){\Text(#1,#2)[l]{{$\scriptscriptstyle #3$}}}
\def\Textsr(#1,#2,#3){\Text(#1,#2)[r]{{$\scriptscriptstyle #3$}}}
\def\Textstl(#1,#2,#3){\Text(#1,#2)[tl]{{$\scriptscriptstyle #3$}}}
\def\Textsbl(#1,#2,#3){\Text(#1,#2)[bl]{{$\scriptscriptstyle #3$}}}
\def\Textstr(#1,#2,#3){\Text(#1,#2)[tr]{{$\scriptscriptstyle #3$}}}
\def\Textsbr(#1,#2,#3){\Text(#1,#2)[br]{{$\scriptscriptstyle #3$}}}
\def\Lwidth{1}

\def\TopoSBnew(#1,#2,#3){\piccc{#1(0,15)(15,15) #2(30,15)(15,0,180)%
 #3(30,15)(15,180,360) #1(45,15)(60,15)%
 \Textl(45,30,\tilde R)\Textl(45,0,\tilde Q+\tilde R)\Textb(3,18,\tilde Q)}}
\def\Bqu(#1,#2)(#3,#4,#5){\SetWidth{2.0}\ArrowArc(#1,#2)(#3,#4,#5)%
\SetWidth{1.0}}
\def\ScatA{\picc{%
 \Laqu(15,30)(30,15)%
 \Lqu(12,15)(30,15)%
 \Laqu(15,0)(30,15)%
 \Lagh(30,15)(45,15)%
 \Textsl(6,21,1)%
 \Textsl(3,10.5,2)%
 \Textsl(6,0,3)%
 \Textsr(40,10.5,Q)%
}}
\def\ScatB{\picc{%
 \Lqu(15,30)(30,15)%
 \Lqu(30,15)(45,30)%
 \Laqu(15,0)(30,15)%
 \Lagh(30,15)(45,0)%
 \Textsl(5,21,2)%
 \Textsr(38,21,1)%
 \Textsl(5,0,3)%
 \Textsr(40,0,Q)%
}}
\def\ScatC{\picc{%
 \Laqu(15,30)(30,15)%
 \Laqu(30,15)(45,30)%
 \Laqu(15,0)(30,15)%
 \Lagh(30,15)(45,0)%
 \Textsl(5,21,1)%
 \Textsr(38,21,2)%
 \Textsl(5,0,3)%
 \Textsr(40,0,Q)%
}}
\def\ScatD{\picc{%
 \Laqu(15,30)(30,15)%
 \Lqu(30,15)(45,30)%
 \Lqu(15,0)(30,15)%
 \Lagh(30,15)(45,0)%
 \Textsl(5,21,1)%
 \Textsr(38,21,3)%
 \Textsl(5,0,2)%
 \Textsr(40,0,Q)%
}}
\def\ScatE{\picc{%
 \Laqu(15,15)(30,15)%
 \Laqu(30,15)(45,30)%
 \Lagh(30,15)(45,15)%
 \Lqu(30,15)(45,0)%
 \Textsl(5,10.5,1)%
 \Textsr(38,21,2)%
 \Textsr(40,10.5,Q)%
 \Textsr(38,0,3)%
}}
\def\ScatF{\picc{%
 \Lqu(15,15)(30,15)%
 \Lqu(30,15)(45,30)%
 \Lagh(30,15)(45,15)%
 \Lqu(30,15)(45,0)%
 \Textsl(5,10.5,2)%
 \Textsr(38,21,1)%
 \Textsr(40,10.5,Q)%
 \Textsr(38,0,3)%
}}
\def\ScatG{\picc{%
 \Laqu(15,15)(30,15)%
 \Lqu(30,15)(45,30)%
 \Lagh(30,15)(45,15)%
 \Laqu(30,15)(45,0)%
 \Textsl(5,10.5,3)%
 \Textsr(38,21,1)%
 \Textsr(40,10.5,Q)%
 \Textsr(38,0,2)%
}}
\def\ScatH{\pic{%
 \Lqu(0,15)(15,30)%
 \Laqu(0,15)(15,20)%
 \Lagh(0,15)(15,10)%
 \Lqu(0,15)(15,0)%
 \Textsr(18,21,1)%
 \Textsr(18,14,2)%
 \Textsr(19,6.5,Q)%
 \Textsr(18,0,3)%
}}
\newcommand{\A}{\hat\nu_\alpha}
\renewcommand{\B}{\,\hat{\!\bar{\nu}}_\beta}
\newcommand{\tss}{\tilde s_0}
\newcommand{\qq}{\tilde q_0}
\newcommand{\rr}{\tilde r_0}
\renewcommand{\tt}{\tilde t_0}
\newcommand{\uu}{\tilde u_0}
\newcommand{\ssfe}{\tilde s_{0\fe}}
\newcommand{\ttfe}{\tilde t_{0\fe}}
\newcommand{\uufe}{\tilde u_{0\fe}}
\newcommand{\rrbo}{\tilde r_{0\bo}}

\makeatletter \@addtoreset{equation}{section} \makeatother
\renewcommand{\theequation}{\arabic{section}.\arabic{equation}}
\makeatletter
\renewcommand\section{\@startsection {section}{1}{\z@}%
                                   {-5.5ex \@plus -1ex \@minus -.2ex}
                                   {2.3ex \@plus.2ex}%
                                   {\normalfont\large\bfseries}}
\renewcommand\subsection{\@startsection{subsection}{2}{\z@}%
                                     {-3.25ex\@plus -1ex \@minus -.2ex}%
                                     {1.5ex \@plus .2ex}%
                                     {\normalfont\normalsize\bfseries}}
\renewcommand\thesection {\@arabic\c@section}
\renewcommand\thesubsection   {\thesection.\@arabic\c@subsection}
\renewcommand{\@seccntformat}[1]{%
\csname the#1\endcsname.\hspace{1.0em}}
\makeatother

\begin{document}

\begin{titlepage}
\begin{flushright}
TU-770\\
BI-TP 2006/17\\
CERN-PH-TH/2006-085\\
hep-ph/0605209\\
\end{flushright}
\begin{centering}
\vfill

{\Large{\bf On the hadronic contribution to sterile neutrino production}}

\vspace{0.8cm}

Takehiko~Asaka$^\rmi{a}$, 
Mikko~Laine$^\rmi{b}$, 
Mikhail~Shaposhnikov$^\rmi{c,d}$ 

\vspace{0.8cm}

$^\rmi{a}${\em
Department of Physics, Tohoku University, 
Sendai 980-8578, Japan\\}

\vspace{0.3cm}

$^\rmi{b}${\em
Faculty of Physics, University of Bielefeld, 
D-33501 Bielefeld, Germany\\}

\vspace{0.3cm}

$^\rmi{c}${\em
Institut de Th\'eorie des Ph\'enom\`enes Physiques, EPFL, 
CH-1015 Lausanne, Switzerland\\}

\vspace{0.3cm}

$^\rmi{d}${\em
Theory Division, CERN, 
CH-1211 Geneva 23, Switzerland\\}

\vspace*{0.8cm}
 
\mbox{\bf Abstract}

\end{centering}

\vspace*{0.3cm}
 
\noindent
Sterile neutrinos with masses in the keV range are considered to be a
viable candidate for warm dark matter.  The rate of their production 
through active-sterile neutrino transitions peaks, however, at
temperatures of the order of the QCD scale,  which makes it difficult
to estimate their relic abundance quantitatively, even if the mass of
the sterile neutrino and its mixing angle were known.  We derive here
a relation, valid to all orders in the strong  coupling constant,
which expresses the production rate  in terms of the spectral
function associated with active neutrinos.  The latter can in turn be
expressed as a certain convolution  of the spectral functions related
to various  mesonic current-current correlation functions,   which
are being actively studied in other physics contexts.  In the naive
weak coupling limit, the appropriate Boltzmann  equations can be
derived from our general formulae.
\vfill
\noindent
 

\vspace*{1cm}
 
\noindent
June 2006

\vfill

\end{titlepage}

%
\section{Introduction}

The problem of explaining the nature of Dark Matter is a central 
one for cosmology. The most popular attempt  is to postulate the
existence of a relatively heavy Cold Dark Matter (CDM) particle
related, perhaps, to softly broken supersymmetry. However, other
particle physics candidates can also be envisaged, and there has
been a recent revival particularly in Warm Dark Matter (WDM) scenarios. 

While WDM is just an alternative to CDM from the Dark
Matter point of view, the issue becomes quite intriguing when other
physics considerations are added to the picture. In particular, WDM
could be realised through the existence of right-handed sterile
neutrinos~\cite{wdm} (see also Ref.~\cite{ot}). This possibility
may then lead to some astrophysical applications~\cite{astro}.
Moreover, if there are three sterile neutrinos in total, one for each 
known generation,  then they can be combined to a minimal theoretical
framework,  dubbed the $\nu$MSM in Ref.~\cite{numsm}, and used to
explain also the known properties of neutrino
oscillations~\cite{numsm} and  baryogenesis~\cite{osc,baryo}.  A
phenomenologically successful implementation can be obtained provided
all  three right-handed neutrinos have masses smaller than  the
electroweak scale.  The role of WDM is played by the lightest 
right-handed neutrino, whereas the two other ones  should have masses
in the range ${\cal O}(1-20)$ GeV and be very  degenerate to produce
the observed baryon asymmetry~\cite{baryo}. In addition, 
an extension of the  $\nu$MSM by a real scalar field, inflaton, 
allows to incorporate inflation~\cite{st}. 

Apart from its mass, $\Ms$, the WDM neutrino is also characterised  
by its mixing angle with active neutrinos, $\theta$.  At present the
strongest observational constraints on $\Ms$ and $\theta$ come
from two sides: structure formation in the form of Lyman-$\alpha$
forest observations imposes a stringent lower limit 
on $\Ms$~\cite{lss,seljak}, while X-ray constraints exclude 
the region of having simultaneously a ``large''  $\Ms$ and
$\theta$~\cite{life}--\cite{hansen}. More precisely, if the
average momentum of sterile neutrinos at temperatures of a few MeV
coincides with that of active neutrinos then, according 
to Ref.~\cite{seljak}, the WDM neutrino cannot be lighter 
than $\Ms\simeq 14$ keV. For masses in this range 
the mixing angle cannot exceed 
$
 \theta \simeq 2.9 \times 10^{-3}\left(1~\mbox{keV}/ \Ms \right)^{5/2}
$~\cite{boyagal}.

In this corner of the parameter space the interactions of sterile
neutrinos with the particles of the Minimal Standard Model (MSM) 
are so weak that the sterile neutrinos cannot equilibrate in 
the Early Universe via active-sterile transitions~\cite{wdm}. 
Therefore, information about initial conditions does {\em not} get lost; 
initial conditions do in general play a role for the current
abundance. Evidently, it would be convenient to fix the initial
conditions at some temperature low enough such that only the 
$\nu$MSM degrees of freedom play a role. Apart from initial conditions, 
one also has to fix the values of all nearly conserved quantum numbers 
in the MSM, such as the lepton and baryon asymmetries. 
Alternatively, one can specify the physics beyond the $\nu$MSM 
and thus determine the initial conditions dynamically;  
an example of such a computation can be found in Ref.~\cite{st}, 
where the main source of sterile neutrino production was 
associated with inflation. 

One can imagine circumstances, however, where the initial
conditions only play a subdominant role. In this case 
the production mechanism of the sterile neutrino WDM can be 
attributed to active-sterile neutrino mixing, as in the
Dodelson-Widrow scenario~\cite{wdm}. The requirements for this
situation can be formulated as follows: suppose that the initial
condition is that the concentration of sterile neutrinos is zero 
(which is possible if the interactions of sterile neutrinos with all
particles beyond the $\nu$MSM such as the inflaton are negligible) 
and that there are no significant lepton asymmetries (lepton
asymmetries corresponding to a chemical potential $\mu/T \gsim 10^{-3}$ 
would play an important role in
sterile neutrino production~\cite{ReSigmamu}). 
Suppose also that the two heavier sterile neutrinos, 
present in the $\nu$MSM, 
are heavy enough so that their decays in the Early Universe do not
produce any entropy~\cite{late}. Then the WDM abundance 
can be expressed as a function of the mass $\Ms$ and the angle 
$\theta$~\cite{wdm,life,oldwdm,oldwdm1}. Matching the
observed abundance one gets a relation between 
$\Ms$ and $\theta$, which can be confronted with the 
observational bounds mentioned above.

Approximate $\Ms$-$\theta$ relations derived along these lines have been 
presented in Refs.~\cite{wdm,life,oldwdm,oldwdm1}. According to the most 
recent analysis~\cite{oldwdm1},
\be
 \theta \approx
 1.3 \times 10^{-4} \left(\frac{1~\mbox{keV}}{\Ms}\right)^{0.8}~,
 \label{angle}
\ee
where a dark matter abundance 
$\Omega_\rmi{DM} \approx 0.22$ and a QCD crossover transition temperature
$T_\rmi{QCD} \approx 170$ MeV have been inserted. 
If true, the combination of the Lyman-$\alpha$
bounds~\cite{seljak} and X-ray bounds~\cite{boyagal} mentioned above
rules out the Dodelson-Widrow scenario. However, other production mechanisms 
such as resonant production due to large lepton 
asymmetries~\cite{ReSigmamu} (see also Ref.~\cite{ak}) 
or due to inflaton interactions~\cite{st} are feasible~\cite{late}.

Due to the importance of the problem our aim here is to reanalyse the
$\Ms$-$\theta$ relation within the Dodelson-Widrow scenario. In fact,
a computation of the sterile neutrino production rate represents a very
non-trivial theoretical challenge. The reason is that the region of
temperatures at which sterile neutrinos are produced most intensely
is \cite{wdm} 
\be
 T\sim 150~\mbox{MeV}
 \left(\frac{\Ms}{1~{\rm keV}}\right)^\frac{1}{3}
 \;. \la{T}
\ee
At higher temperatures their production is suppressed because of 
medium effects \cite{bd}. The temperature in \eq\nr{T} is very close
to the pseudocritical temperature of the QCD crossover. Therefore, 
neither the dilute hadronic gas approximation nor the 
weakly interacting quark-gluon plasma picture is expected to 
provide for an accurate description. 

The presence of strongly interacting hadrons at these temperatures
leads to two sources of uncertainties. A well-known one is related to
the hadronic equation of state, needed for the time-temperature
relation in the expanding Universe, entering the sterile neutrino
production equation. Unfortunately, experiments with heavy ion
collisions cannot directly measure the equation of state
of hadronic matter. In addition, present lattice simulations 
with light dynamical quarks involve uncontrolled systematic 
uncertainties, such as the absence of a 
continuum limit extrapolation.  In other words, the equation of 
state of QCD is  only known  
approximately  in this temperature range and contains significant systematic 
uncertainties (for a recent discussion, see Ref.~\cite{phen}).

At the same time, the equation of state does matter in the computation of 
the sterile neutrino abundance. As has been mentioned already 
in Ref.~\cite{wdm} and elucidated further in Refs.~\cite{oldwdm},
even the purely leptonic contribution to the abundance depends
significantly on the effective number of massless
degrees of freedom, $g_*$, which in turn changes dramatically, 
from about $60$ down to about $20$, 
when the quark-gluon plasma cools to a hadronic gas.
However, neither the uncertainties of 
the equation of state,
nor the subsequent uncertainties in the 
$\Ms$-$\theta$ relation, have been 
exhaustively investigated in these works.

There is also a second type of a hadronic uncertainty,
which has a dynamical character. Sterile
neutrinos can be produced in reactions of two types, the first 
containing leptons only and the second having hadrons in the initial
state. The hadronic reactions were omitted in Refs.~\cite{wdm,life}.
Processes with quarks were mentioned in Ref.~\cite{oldwdm} (not in
Ref.~\cite{oldwdm1}), but without an explanation of how they were treated
in the QCD crossover region. 

It is this second uncertainty that 
is the focus of the present paper. Our goal is to 
set up a general formalism for attacking it. In a later work, a numerical
analysis of the $\Ms$-$\theta$ relation and its uncertainties will be
presented. More concretely,  the current number
density of the WDM neutrinos in the Dodelson-Widrow scenario 
is given by $\theta^2 F + {\cal O}(\theta^4)$.  
In this paper we derive an expression which allows 
to relate the  coefficient $F$ to a certain equilibrium Green's
function,  the so-called active neutrino spectral function,  defined
within the MSM, while all dependence on the parameters of the
$\nu$MSM appears as the prefactor $\theta^2$.  

It is appropriate to stress that analogous relations exist in other
contexts as well. For instance, it can be shown  that the photon
spectral function computed within the MSM  determines the production
rate of on-shell photons and  dilepton pairs from strongly
interacting systems such as  colliding heavy nuclei~\cite{dilepton},
and the production rate of active neutrino pairs from  hot/dense
astrophysical environments such as the cores of neutron
stars~\cite{nunu}.  Nevertheless, we are not aware of the existence
of such relations  in the present context, so we want to discuss 
their derivation in a hopefully pedagogic manner. 

Given the Green's function, it should still be evaluated.  As we have
already mentioned, this turns out to be a very difficult task, since
the sterile neutrino production rate peaks at  temperatures of the
order of the QCD scale.  In this temperature range strong
interactions play  a dominant role, and perturbative methods fail. In
the second part of our paper, we thus  show how the active neutrino
spectral function can be related to various  vector and axial-vector
current-current correlation functions  defined within high
temperature QCD. Such objects have previously been studied  with a
variety of methods, such as chiral perturbation theory, QCD sum
rules, lattice QCD, and resummed weak-coupling perturbation theory, 
and also possess independent physics  applications, particularly in
connection with the photon and dilepton pair production rate
computations mentioned above. 

The plan of this paper is the following.  In \se\ref{se:master} we
derive the expression alluded to above,  expressing the sterile
neutrino production rate in terms of the  spectral function of active
neutrinos, computable within the MSM.  In \se\ref{se:hadron} we
relate the hadronic contribution to the active neutrino spectral
function to mesonic current-current correlation  functions, which can
be defined within QCD. We also show how the result reduces to 
certain Boltzmann equations in the naive (unresummed) weak-coupling
limit.  We conclude and outline future prospects in
\se\ref{se:concl}. The three appendices contain certain basic
definitions for the various bosonic and fermionic Green's functions
that appear in our study,  and an alternative derivation for
\se\ref{se:master}. 

%
\section{General formula for the sterile neutrino production rate}
\la{se:master}

%
\subsection{Notation}

It is a matter of convention whether the right-handed neutrinos are
represented as Weyl, right-handed Dirac, or Majorana fermions.
Choosing here the last option, the Minkowskian Lagrangian of
$\nu$MSM  can be written as
\be
 \mathcal{L} = 
 \fr12 \bar{\tilde N_I} i \msl{\partial} \tilde N_I 
 - \fr12 M_I \bar{\tilde N_I} \tilde N_I
 - F_{\alpha I} \bar L_\alpha \tilde \phi\, a_R \tilde N_I
 - F_{\alpha I}^* \bar{\tilde N_I} \tilde \phi^\dagger a_L L_\alpha
 + \mathcal{L}_\rmi{MSM}
 \;,  \la{LM}  
\ee
where $\tilde N_I$ are Majorana spinors, repeated indices are  summed
over,  $M_I$ are Majorana masses that we have chosen to be  real in
this basis, $L_\alpha$ are the weak interaction eigenstates of the
active lepton doublets, $F_{\alpha I}$ are elements of a  3$\times$3
complex Yukawa matrix,  $\tilde\phi = i \tau_2 \phi^*$ is the
conjugate  Higgs doublet, and $a_L \equiv (1-\gamma_5)/2$,  $a_R
\equiv (1+\gamma_5)/2$ are chiral projectors.

After electroweak symmetry breaking,  $\langle \tilde \phi \rangle =
(v/\sqrt{2},0)$ ,  we define the matrix
\be
 (M_D)_{\alpha I} \equiv \frac{v F_{\alpha I}}{\sqrt{2}}
 \;, \la{MD}
\ee
where $v\simeq 246$~GeV.
The various neutrino fields, active and sterile, then couple
to each other, and diagonalising the mass matrix we can define
mass eigenstates in the usual way. However, a sterile neutrino
of type $I$ couples to an active neutrino mass eigenstate of 
type $a$ with a very small angle only, 
\be
 \theta_{Ia} \simeq  \sum_{\alpha=1}^3 
 \frac{(M_D^\dagger)_{I\alpha}}{M_I} U_{\alpha a}
 \;,
\ee
where $U_{\alpha a}$ is the mixing matrix between the active neutrino
interaction and mass bases. As mentioned above,   the
phenomenologically relevant part of the parameter space corresponds
to values $\theta \ll 1$. Therefore,  the sterile neutrino
interaction eigenstate of type $I$ is to a very good approximation
also  a sterile neutrino mass eigenstate,  and we can ignore the
distinction in the following. To fix the conventions, $I=1$
corresponds to the lightest sterile neutrino, contributing to 
warm dark matter.

The Green's functions that we will need
involve a lot of sign and other conventions whose definitions
are unfortunately not unique in the literature. We therefore
explicitly state our conventions in Appendices A and B, 
for Green's functions made out of bosonic and fermionic 
operators, respectively. Our metric convention is ($+$$-$$-$$-$).

%
\subsection{Derivation of the master equation}

According to our assumption, the concentration of sterile neutrinos 
was zero at very high temperatures, $T \gg 1$ GeV. Moreover, because of the
smallness of its Yukawa coupling, the lightest sterile neutrino
never equilibrated. In this section we show that these two facts
allow us to express the production rate of this neutrino through 
a certain well-defined {\it equilibrium} Green's function within 
the MSM. The consideration below is very 
general and uses only the basic principles of thermodynamics 
and quantum field theory. In particular, it does not require any 
solution of kinetic equations, nor a discussion of coherence or 
its loss due to collisions, or the like.

The general way we proceed with the derivation is equivalent
to how fluctuation-dissipation relations, or linear response
formulae, are usually derived (see, e.g., Refs.~\cite{old,mlb}). 
There exists, however, also an alternative
derivation, which makes more direct contact with particle states
and the related transition matrix elements and which is also 
somewhat shorter. The price is that this derivation appears
to be slightly less rigorous. Nevertheless, the end result is 
identical, so we present the alternative derivation in Appendix C.

We disregard first the Universe expansion, which can be added
later on (cf.\ \eq\nr{expansion}). 
Let $\hat\rho$ be the density matrix for $\nu$MSM,
incorporating all degrees of freedom, and $\hat H$ the corresponding
full Hamiltonian operator. Then the equation for the density matrix is 
\be 
 i \frac{{\rm d} \hat\rho(t)}{{\rm d} t} =[\hat H,\hat\rho(t)]
 \;.
 \label{liuv}
\ee
We now split $\hat H$ in the form
\be
 \hat H= \hat H_\rmi{MSM} + \hat H_\rmi{S} + \hat H_\rmi{int}
 \;,
\ee
where $\hat H_\rmi{MSM}$ is the complete Hamiltonian of the
MSM,  $\hat H_\rmi{S}$ is the free Hamiltonian of sterile
neutrinos,  and $\hat H_\rmi{int}$, which is proportional to the
sterile neutrino Yukawa couplings, contains the interactions between
sterile neutrinos and the particles of the MSM. To find the
concentration of sterile neutrinos, one has to solve \eq\nr{liuv}
with some initial condition. Following~\cite{wdm}, we will {\em
assume} that the initial concentration of sterile neutrinos is zero,
that is
\be
 \hat \rho (0) = \hat \rho_\rmi{MSM}\otimes |0\rangle\langle 0|
 \;, 
 \label{in}
\ee
where  
$
 \hat \rho_\rmi{MSM} 
 = Z^{-1}_\rmi{MSM} \exp(-\beta \hat H_\rmi{MSM})
$, 
$
 \beta \equiv 1/T 
$,  
is the equilibrium MSM density matrix at a temperature $T$,  and
$|0\rangle$ is the vacuum state for sterile neutrinos.  The physical
meaning of \eq\nr{in} is clear:  it describes a system with no
sterile neutrinos, while all MSM particles  are in thermal
equilibrium.

Considering now $\hat H_0=\hat H_\rmi{MSM} + \hat H_\rmi{S}$  as a
``free'' Hamiltonian, and $\hat H_\rmi{int}$ as an  interaction
term, one can derive an equation for the density  matrix in the
interaction picture, 
$ 
 \hat \rho_\rmi{I} \equiv \exp(i \hat H_0 t)\hat \rho \exp(-i \hat H_0 t)
$, 
in the standard way:
\be 
 i \frac{{\rm d} \hat \rho_\rmi{I}(t)}{{\rm d} t} =
 [\hat H_\rmi{I}(t),\hat\rho_\rmi{I}(t)]
 \;. \label{liuv1}
\ee
Here, as usual, 
$
 \hat H_\rmi{I}= \exp(i \hat H_0 t) \hat H_\rmi{int} \exp(-i \hat H_0 t)
$ is
the interaction Hamiltonian in the interaction picture. Now, 
perturbation theory with respect to $\hat H_\rmi{I}$ can be used 
to compute the time evolution of $\hat\rho_\rmi{I}$; 
the first two terms read
\ba
 \hat \rho_\rmi{I}(t) = \hat\rho_0
 - i \int_0^t \! {\rm d} t' \, 
 [\hat H_\rmi{I}(t'), \hat\rho_0]
 + (-i)^2 
 \int_0^t \! {\rm d} t' \,
 \int_0^{t'} \! {\rm d} t'' \,
 [\hat H_\rmi{I}(t'),[\hat H_\rmi{I}(t''), \hat\rho_0]]
 + ... \;,
 \label{pert}
\ea
where $\hat\rho_0 \equiv \hat\rho(0) = \hat\rho_\rmi{I}(0)$.
Note that perturbation theory with $\hat H_\rmi{I}$ breaks down at 
a certain time $t \simeq t_\rmi{eq}$ due to so-called secular terms. 
After $t_\rmi{eq}$ sterile neutrinos enter thermal equilibrium and 
their concentration needs to be computed by other means. For
us $t \ll t_\rmi{eq}$ and perturbation theory works well.

We are interested in the distribution function of the sterile
neutrinos. It is associated with the operator
\be
 \frac{{\rm d} \hat N_{I}}{{\rm d}^3\vec{x}\, {\rm d}^3\vec{q}}
 \equiv \frac{1}{V} \sum_{s=\pm 1} \hat a^\dagger_{I;\vec{q},s} 
                      \hat a^{\mbox{ }}_{I;\vec{q},s}
 \;, 
\ee
where 
$
 \hat a^\dagger_{I;\vec{q},s} 
$ 
is the creation operator 
of a sterile neutrino of type $I$, momentum $\vec{q}$, and spin state $s$, 
normalised as 
\be
 \{ \hat a^{\mbox{ }}_{I;\vec{p},s} , \hat a^\dagger_{J;\vec{q},t} \}
 = \delta^{(3)}(\vec{p}-\vec{q})\delta_{I\!J}\delta_{st}
 \;, \la{norm}
\ee
and $V$ is the volume of the system.
Then the distribution function 
$
 {{\rm d} N_{I}} / {{\rm d}^3\vec{x}\, {\rm d}^3\vec{q}}
$
(number of sterile neutrinos of type $I$ per 
${{\rm d}^3\vec{x}\, {\rm d}^3\vec{q}}$) is given by
\be
 \frac{{\rm d} N_{I}(x,\vec{q})}{{\rm d}^3\vec{x}\, {\rm d}^3\vec{q}}
 = 
 \tr \biggl[ 
  \frac{{\rm d} \hat N_{I}}{{\rm d}^3\vec{x}\, {\rm d}^3\vec{q}}
  \hat\rho_\rmi{I}(t)
 \biggr]
 \;. \label{Np}
\ee
One can easily see that the first term in \eq\nr{pert} 
does not contribute in \eq\nr{Np} since $\hat H_\rmi{I}$ 
is linear in $\hat a^\dagger_{I;\vec{q},s}$ and  
$\hat a^{\mbox{ }}_{I;\vec{q},s}$.
Thus, we get that to $\rmO(\theta^2)$
the rate of sterile neutrino production reads
\be
 \frac{{\rm d} N_{I}(x,\vec{q})}{{\rm d}^4{x}\, {\rm d}^3\vec{q}}
 = - \frac{1}{V} 
 \tr \biggl\{ 
 \sum_{s=\pm1} \hat a^\dagger_{I;\vec{q},s} \hat a^{\mbox{ }}_{I;\vec{q},s}
 \int_0^t \! {\rm d} t' \,
 [\hat H_\rmi{I}(t),[\hat H_\rmi{I}(t'), \hat\rho_0]]
 \biggr\}
 \;. \label{rate}
\ee

For small temperatures $T \ll M_W$, 
the Higgs field in $\hat H_\rmi{I}$ can safely 
be replaced through its vacuum expectation value, 
so that \eqs\nr{LM}, \nr{MD} imply
\be
 \hat H_\rmi{I} = 
 \int\! {\rm d}^3 \vec{x} \, 
 \Bigl[
  (M_D)_{\alpha I} \,\hat{\!{\bar\nu}}_\alpha a_R \hat N_I + 
  (M_D^*)_{\alpha I} \,\hat{\!\bar N}_I a_L \hat \nu_\alpha
 \Bigr]
 \;, \la{HI}
\ee
where now $\hat N_I$ is a Majorana spinor field operator.
The $\hat N_I$ can be treated
as free on-shell field operators and can hence be written as 
\ba
 \hat{N}_I(x) & = & 
 \int\! \frac{{\rm d}^3 \vec{p}}{\sqrt{\raise-0.1ex\hbox{$(2\pi)^3 2 p^0$}}}
 \sum_{s=\pm 1}
 \Bigl[ 
  \hat{a}^{\mbox{ }}_{I;\vec{p},s} 
   u(I;\vec{p},s) e^{-iP\cdot x} + 
  \hat{a}^{\dagger}_{I;\vec{p},s} 
   v(I;\vec{p},s) e^{iP\cdot x}   
 \Bigr]
 \;, \la{norm1} \\ 
 \,\hat{\!\bar{N}}_I(x) & = & 
 \int\! \frac{{\rm d}^3 \vec{p}}{\sqrt{\raise-0.1ex\hbox{$(2\pi)^3 2 p^0$}}}
 \sum_{s=\pm 1}
 \Bigl[ 
  \hat{a}^{\dagger}_{I;\vec{p},s} \bar u(I;\vec{p},s) e^{iP\cdot x} + 
  \hat{a}^{\mbox{ }}_{I;\vec{p},s} \bar v(I;\vec{p},s) e^{-iP\cdot x}   
 \Bigr]
 \;, \la{norm2}
\ea
where we assumed the normalization in \eq\nr{norm}, 
and $p^0\equiv E_\vec{p}^{(I)} \equiv \sqrt{\vec{p}^2 + M_I^2}$, 
$P \equiv (p^0,\vec{p})$.
The spinors $u,v$ satisfy the completeness relations 
$
 \sum_s u(I;\vec{p},s) \bar u(I;\vec{p},s) = \slash \!\!\!{p}\,\,\, + M_I
$, 
$
 \sum_s v(I;\vec{p},s) \bar v(I;\vec{p},s) = \slash \!\!\!{p}\,\,\, - M_I
$, 
and their Majorana character requires that 
$
 u = C \bar{v}^T
$, 
$
 v = C \bar{u}^T
$,
where $C$ is the charge conjugation matrix. Inserting the free
field operators into \eq\nr{HI}, we can rewrite it as 
\be
 \hat H_\rmi{I} = 
 \int\! {\rm d}^3 \vec{x}  
 \int\! \frac{{\rm d}^3 \vec{p}}{\sqrt{\raise-0.1ex\hbox{$(2\pi)^3 2 p^0$}}}
 \sum_{s=\pm 1}
 \biggl[ 
   \hat a^\dagger_{I;\vec{p},s} \, \hat J_{I;\vec{p},s}(x) \, e^{i P\cdot x} + 
   \hat J^\dagger_{I;\vec{p},s}(x) \, \hat a_{I;\vec{p},s} \, e^{-i P\cdot x}
 \biggr] 
  \;, \la{HI2}
\ee
where 
\be
 \hat J_{I;\vec{p},s}(x) \equiv  
  - (M_D)_{\alpha I} \,\hat{\!{\bar\nu}}_\alpha(x) a_R v(I;\vec{p},s) + 
  (M_D^*)_{\alpha I} \, \bar{u}(I;\vec{p},s) a_L \hat \nu_\alpha(x)
 \;. \la{J}
\ee

It remains to take the following steps:
\begin{itemize}
\item[(i)]
We insert \eq\nr{HI2} into \eq\nr{rate} and remove the 
sterile neutrino creation and annihilation operators, 
by making use of \eq\nr{norm}.

\la{steps}
\item[(ii)]
This leaves us with various types of two-point correlators
of the active neutrino field operators. We now note that
correlators of the type 
$
 \langle \B(x') \,\hat{\!\bar{\nu}}_\alpha(x) \rangle
$
and 
$
 \langle \hat\nu_\beta(x') \A(x) \rangle
$,
where $\langle ... \rangle \equiv \tr[\hat\rho_\rmi{MSM}(...)]$
and we have generalised the notation so that $\alpha, \beta$
incorporate also the Dirac indices, vanish, since lepton 
numbers are conserved within the MSM.

\item[(iii)]
The non-vanishing two-point functions contain the spinors 
$u,v$ in a form where the standard completeness relations
mentioned above can be used. The mass terms $M_I$ that are 
induced this way get projected out by $a_L,a_R$.

\item[(iv)]
Introducing the notation in \eqs\nr{fL}, \nr{fS}, the remaining
two-point correlator can be written as
\be
 \langle \A(x')\B(x)+\B(x')\A(x) \rangle 
 = 
 \int\! \frac{{\rm d}^4 P}{(2\pi)^4}
 e^{-i P\cdot(x-x')}
 \Bigl[ 
  \Pi^{>}_{\alpha\beta}(-P) -  
  \Pi^{<}_{\alpha\beta}(P)
 \Bigr]
 \;. \la{iv}
\ee
There is another term with the same structure 
but with $x\leftrightarrow x'$.

\item[(v)]
It remains to carry out the integrals over the space and time 
coordinates. Taking the limit $t\to\infty$, they yield 
\be
 \lim_{t\to\infty}
 \int\! {\rm d}^3 \vec{x} \,
 \int\! {\rm d}^3 \vec{x}' \,
 \int_0^t \! {\rm d}t'\,
 \Bigl[ 
  e^{i(Q-P)\cdot(x-x')} + 
  e^{i(P-Q)\cdot(x-x')}
 \Bigr] = V (2\pi)^4 \delta^{(4)}(P-Q)
 \;,
\ee
which allows to
cancel $1/V$ from \eq\nr{rate}
and remove $P$-integration from \eq\nr{iv}.

\end{itemize}

As a result of all these steps, we obtain 
\be
 \frac{{\rm d} N_I(x,\vec{q})}{{\rm d}^4 x\,{\rm d}^3 \vec{q}}
 = \frac{1}{(2\pi)^3 2 q^0}
   (M_D^*)_{\alpha I} (M_D)_{\beta I}
 \tr\Bigl\{ 
  \bsl{Q}
  a_L 
   \Bigl[
    \Pi^{>}_{\alpha\beta}(-Q) - \Pi^{<}_{\alpha\beta}(Q) 
   \Bigr]
  a_R
 \Bigr\}
 \;, \la{raw}
\ee
where we have returned to the convention that $\alpha, \beta$
label generations, and have expressed the Dirac part through
a trace. Inserting \eq\nr{fLSrel}; making use of the fact that 
$
 1-\nF{}(-q^0) = \nF{}(q^0)
$, where 
$\nF{}(x)\equiv 1/[\exp(\beta x) + 1]$;
and observing that lepton generation conservation within the MSM 
restricts the indices $\alpha, \beta$ to be equal, we finally obtain 
the master relation
\be
 \frac{{\rm d} N_I(x,\vec{q})}{{\rm d}^4 x\,{\rm d}^3 \vec{q}}
 =  R(T,\vec{q})
 \equiv \frac{2 \nF{}(q^0)}{(2\pi)^3 2 q^0}
   \sum_{\alpha = 1}^{3}  
   |M_D|^2_{\alpha I} 
  \tr\Bigl\{ 
  \bsl{Q}
  a_L 
   \Bigl[
    \rho_{\alpha\alpha}(-Q) + \rho_{\alpha\alpha}(Q) 
   \Bigr]
  a_R
 \Bigr\} 
 \;, \la{master}
\ee
where $\rho$ is called the spectral function (\eq\nr{frho}). We
stress again that this relation is valid only provided that  the
number density of sterile neutrinos created is much smaller than  their
equilibrium concentration, which however is always the case in the
phenomenologically interesting part of the parameter space, at least
for $I=1$.

In an expanding Universe, with a Hubble rate $H$, 
the physical momenta redshift as $\vec{q}(t) = \vec{q}(t_0)\, a(t_0)/a(t)$, 
where $a(t)$ is the scale factor. This implies that the time
derivative gets replaced with 
$
 {\rm d}/{\rm d}t = \partial/\partial t - H q_i \partial/\partial q_i
$~\cite{expansion}, 
and \eq\nr{master} becomes
\be
  \biggl[ \frac{\partial}{\partial t} - 
  H q_i \frac{\partial}{\partial q_i}\biggr]  
  \frac{{\rm d} N_I(x,\vec{q})}{{\rm d}^3 \vec{x}\,{\rm d}^3 \vec{q}}
  = R(T,\vec{q})
 \;. \la{expansion}
\ee

%
\section{Hadronic contribution to the active neutrino spectral function}
\la{se:hadron}

%
\subsection{Notation}

As stated by \eq\nr{master}, we need to estimate the active 
neutrino spectral function within the MSM. Given that higher-order
corrections can be important, this task has to be consistently 
formulated within finite-temperature field theory. There are 
in principle two ways to go forward, the 
real-time and the imaginary-time formalisms. 
We follow here the latter since it can be set up
also beyond perturbation theory.

Within the imaginary-time formalism, the spectral function
can be obtained through a certain analytic continuation
of the Euclidean active neutrino propagator. With the 
conventions specified in Appendix~B, we denote the Euclidean
propagator by $\Pi^E_{\alpha\beta}(\tilde q_0, \vec{q})$
(cf.\ \eq\nr{fE}). Carrying out an analytic continuation, 
we define 
\be
 \Pi^E_{\alpha\beta}(-i[q^0\pm i 0^+],\vec{q}) \equiv
 \re \Pi^R_{\alpha\beta}(q^0,\vec{q}) 
 \pm i \im \Pi^R_{\alpha\beta}(q^0,\vec{q})
 \;,  \la{defim}
\ee
where $\Pi^R_{\alpha\beta}$ is called the retarded Green's
function (cf.\ \eq\nr{fR}). The relation shown in \eq\nr{defim} follows
from the spectral representations in \eqs\nr{fRrhorel}, \nr{fErhorel}.
Note that the imaginary part in \eq\nr{defim} 
is {\it defined} as the discontinuity across the real axis:
\ba
 \im \Pi^R_{\alpha\beta}(q^0,\vec{q}) & \equiv & 
 \frac{1}{2i} \disc \Pi^E_{\alpha\beta}(-iq^0,\vec{q})
 \la{sfdef} \\
 & \equiv &   
 \frac{1}{2i} \Bigl[ 
 \Pi^E_{\alpha\beta}(-i[q^0+i 0^+],\vec{q}) - 
 \Pi^E_{\alpha\beta}(-i[q^0-i 0^+],\vec{q})
 \Bigr] \la{Discdef} \\ 
 & = & 
 \rho_{\alpha\beta}(q^0,\vec{q})
 \;,  \la{rhodef}
\ea
where the last step introduced the 
spectral function (\eq\nr{frho}) 
and made use of \eq\nr{ffinal}.
As the imaginary-time neutrino propagator is time-ordered
by construction (cf.\ \eq\nr{fE}), we can 
use functional integrals for its determination, whereby 
operator labels can be dropped from the fields from now on.

In order to compute the Euclidean propagator, from which
the spectral function follows through \eqs\nr{defim}, \nr{rhodef},
we need to define the Euclidean theory. Given that we are interested
in low temperatures, we can work within
the Fermi-model. The interactions of the active neutrinos
with the hadronic degrees of freedom on which we concentrate 
in this paper, are then contained in the Lagrangian
\ba
 \mathcal{L}_E & = & 2\sqrt{2} G_F 
 \biggl( 
   \bar \nu_\alpha \tilde\gamma_\mu a_L l_\alpha \, \tilde H_\mu^W + 
   \tilde H_\mu^{W\dagger}\, \bar l_\alpha \tilde\gamma_\mu a_L \nu_\alpha 
 + \fr12 \bar \nu_\alpha \tilde\gamma_\mu a_L \nu_\alpha\, \tilde H_\mu^Z 
 \biggr)
 \;, \\ 
 \tilde H_\mu^W & = & 
 \bar d_{\beta B} ' \, \tilde\gamma_\mu a_L \, u_{\beta B} 
 \;, \quad
 \tilde H_\mu^{W\dagger} \; = \;
 \bar u_{\beta B} \, \tilde\gamma_\mu a_L \, d'_{\beta B} 
 \;, \la{HW}
 \\
 \tilde H_\mu^Z & = & 
  \bar{u}_{\beta B} \, \tilde\gamma_\mu 
    \biggl(
       \fr12 - \frac{4x_\rmii{W}}{3} - \frac{\gamma_5}{2}  
    \biggr) u_{\beta B}
  + 
  \bar{d}_{\beta B} \, \tilde\gamma_\mu 
    \biggl(
      -\fr12 + \frac{2 x_\rmii{W}}{3} + \frac{\gamma_5}{2}
    \biggr) d_{\beta B}
 \;, \la{HZ}
\ea
where $\alpha,\beta$ are generation indices, 
$B$ is a colour index,  
$x_\rmii{W} \equiv \sin^2 \theta_\rmii{W}$, and
the Fermi-constant reads $G_F = g_w^2/4\sqrt{2} m_W^2$. 
All fermions are Dirac fields. The fields $d'_{\beta B}$
are related to the mass eigenstates $d_{\beta B}$
with the usual CKM matrix. The Euclidean $\gamma$-matrices 
are defined by $\tilde\gamma_0 \equiv \gamma^0$, 
$\tilde\gamma_i \equiv - i \gamma^i$, and satisfy 
$\tilde\gamma_\mu^{\dagger} = \tilde\gamma_\mu$, 
$\{\tilde\gamma_\mu,\tilde\gamma_\nu\} = 2 \delta_{\mu\nu}$.
We have defined 
$\gamma_5 = \tilde\gamma_0\tilde\gamma_1\tilde\gamma_2\tilde\gamma_3 = 
i \gamma^0\gamma^1\gamma^2\gamma^3$. 
Repeated $\mu$-indices are summed over, and that they
are both down reminds us of the fact 
that we are in the Euclidean space-time.
We also denote 
$\,\slash\!\!\!\!\tilde Q \,\,\,\, \equiv \tilde q_\mu \tilde\gamma_\mu$
and note that if we carry out
the Wick-rotation $\qq \to -i q^0$ 
(cf.\ \eq\nr{defim}) and  simultanously decide
to express the result in terms of Minkowskian rather than
Euclidean Dirac-matrices, then
\be
 i \,\slash \!\!\!\! \tilde Q \,\,\,\, \to 
 \,\slash \!\!\!\! Q \,\,\,\, \equiv q_\mu \gamma^\mu 
 \;. \la{EtoM}
\ee

%
\subsection{General structure of the active neutrino spectral function}

Suppose now that we compute the full Euclidean neutrino self-energy within 
the MSM. Since only left-handed neutrinos experience 
interactions in the MSM, we expect the corresponding Euclidean
action to have the structure 
\be
 S_E = 
 \Tint{\tilde Q_\fe}
 \sum_{\alpha = 1}^{3} \bar\nu_\alpha(\tilde Q)
 a_R [ i \bsl{\tilde Q} + i \bsl{\tilde\Sigma}_{\alpha\alpha}(\tilde Q) ] a_L
 \nu_\alpha(\tilde Q)
 \;, 
\ee
where we have defined the Fourier transforms as 
$
 \nu_\alpha(\tilde x)= \int_{\tilde Q} \exp(i\tilde Q\cdot\tilde x)
 \nu_\alpha(\tilde Q)
$, 
$
 \bar \nu_\alpha(\tilde x)= \int_{\tilde Q} \exp(-i\tilde Q\cdot\tilde x)
 \bar \nu_\alpha(\tilde Q)
$.
To keep the future expressions more compact, we have also
factored the chiral projectors outside of $\tilde\Sigma$, but 
their existence should be kept in mind in the following. With 
the conventions of \eq\nr{fE}, this leads to the Euclidean propagator
\be
 \Pi^E_{\alpha\alpha}(\tilde Q) = a_L 
 \frac{1}{- i \bsl{\tilde Q} + i \bsl{\tilde\Sigma}
 ( -\tilde Q)} a_R 
 = a_L\, \frac{i \bsl{\tilde Q} + i \bsl{\tilde\Sigma}(\tilde Q)}
   {[\tilde Q + \tilde \Sigma(\tilde Q)]^2}\, a_R
 \;, \la{prop}
\ee
where we have made use of the property 
$\bsl{\tilde\Sigma}( -\tilde Q) = - \bsl{\tilde\Sigma}( \tilde Q)$,
following from hermiticity (or, to be more precise, the so-called 
$\gamma_5$-hermiticity that replaces hermiticity in the Euclidean theory: 
the Dirac operator $D$ satisfies $\gamma_5 D^\dagger \gamma_5 = D$)
and CP-invariance. 
We have also left out the flavour indices from $\tilde\Sigma$ 
to compactify the notation somewhat. 

Defining now, in analogy with the Wick-rotation of $\tilde Q$, 
a four-vector $\Sigma_{\mu}\equiv (i\tilde \Sigma_0, \tilde \Sigma_i)$; 
recalling that $\tilde Q^2 = \tilde q_\mu\tilde q_\mu = - Q^2$; 
and making use of \eq\nr{EtoM}, 
we can write a general analytic continuation 
of $\Pi^E_{\alpha\alpha}(\tilde Q)$ as
\be
 \Pi^R_{\alpha\alpha}(q^0,\vec{q}) = 
 \Pi^E_{\alpha\alpha}(-i q^0,\vec{q}) = a_L \frac{-\bsl{Q} - \bsl{\Sigma}(Q)}
 {Q^2 + 2 Q\cdot \Sigma(Q) + \Sigma^2(Q)} a_R
 \;. \la{retd}
\ee
Writing finally
$\Sigma(q^0\pm i 0^+,\vec{q}) \equiv 
\re \Sigma(q^0,\vec{q}) \pm i \im \Sigma(q^0,\vec{q})$ in analogy
with \eq\nr{defim}, making use of \eq\nr{rhodef}, and
employing the symmetry properties 
$\re\Sigma(-Q) = - \re\Sigma(Q)$, 
$\im\Sigma(-Q) =   \im\Sigma(Q)$, 
the master relation of \eq\nr{master} becomes 
\ba
 && \hspace*{-1cm}
 \frac{{\rm d} N_I(x,\vec{q})}{{\rm d}^4 x\,{\rm d}^3 \vec{q}}
  =  \frac{4 \nF{}(q^0)}{(2\pi)^3 2 q^0}
   \sum_{\alpha = 1}^{3}  
   \frac{|M_D|^2_{\alpha I}}
   {\{[Q + \re\Sigma]^2 - [\im \Sigma]^2 \}^2 + 
       4 \{[Q + \re\Sigma]\cdot \im\Sigma \}^2 } \times
 \la{master2} \\ 
 &&  \times   
 \tr \Bigl\{
  \bsl{Q} a_L \Bigl(
    2 [Q + \re\Sigma]\cdot\im\Sigma 
      [ \bsl{Q} + \re \bsl{\Sigma} ]
    - \{
         [Q + \re \Sigma]^2 - [\im\Sigma]^2 
      \} \im\bsl{\Sigma} 
  \Bigr) a_R
  \Bigr\}
  \;, \nonumber
\ea 
where $\Sigma \equiv \Sigma_{\alpha\alpha}(Q)$, and $Q^2 = M_I^2$.

We remark that 
the trace over Dirac matrices 
on the latter row of \eq\nr{master2}
could  trivially be carried out.
Given that this does not simplify the structure in an essential way,
however, we do not write down the corresponding formula explicitly. 
We also note that $\bsl{\Sigma}$ can contain two types of Lorentz 
structures, 
$
 \bsl{\Sigma}(Q) = \bsl{Q} f_1(Q^2,Q\cdot u) + \msl{u} f_2(Q^2,Q\cdot u)
$,
where $u=(1,\vec{0})$ is the plasma four-velocity~\cite{weldon}. However, 
we do not need to make a distinction between these two structures here.

Now, in the absence of (leptonic) chemical 
potentials~\cite{ReSigmamu}, it is easy to see that $\re\Sigma$ 
gets no contributions at $\rmO(g_w^2/m_W^2)$, 
corresponding to 1-loop level within the Fermi model. 
The dominant contributions are $\rmO(g_w^2/m_W^4)$, and 
originate from 1-loop graphs within the electroweak 
theory~\cite{ReSigmaold,ReSigma}. 
The dominant contributions within the Fermi-model are of 2-loop
order, $\rmO(g_w^4/m_W^4)$, and thus suppressed 
with respect to the 1-loop effects from the electroweak theory. 
In contrast, $\im\Sigma$ requires on-shell particles  on the 
inner lines, and cannot at low energies $E \ll m_W$ get 
generated within 1-loop level in the electroweak theory
(more precisely, $\im\Sigma$ is
exponentially suppressed by $\sim\exp(-m_W/T)$). 
The dominant contributions 
are $\rmO(g_w^4/m_W^4)$ and can be computed within the Fermi-model.
It is these contributions that we concentrate on in the following. 
 
For general orientation, it is useful to note that if we assume
$\re\Sigma, \im\Sigma \ll Q$, as is parametrically the case at 
low energies, then \eq\nr{master2} can be simplified to 
\be
  \frac{{\rm d} N_I(x,\vec{q})}{{\rm d}^4 x\,{\rm d}^3 \vec{q}}
 \approx  \frac{4 \nF{}(q^0)}{(2\pi)^3 2 q^0}
   \sum_{\alpha = 1}^{3}  
   \frac{|M_D|^2_{\alpha I}}{Q^2}
   \tr\Bigl[\bsl{Q} a_L \im \bsl{\Sigma} a_R\Bigr]
   \;. \la{master_simple}
\ee
Given that the large-time decay of the retarded propagator $\Pi^R$
is determined by the structure $q^0 + i \im\Sigma^0$ 
(cf.\ \eq\nr{retd}), and given our conventions
in \eq\nr{fR}, we expect the behaviour 
$\Pi^R(x^0)\sim \int_{q^0} \exp(-i q^0 x^0)/(q^0 + i \im\Sigma^0)
\sim \exp(-x^0 \im\Sigma^0)$. Therefore   
$\im\Sigma^0$ has to be {\it positive}; in fact, we can define 
$\im\Sigma^0 = \Gamma_\nu/2$, where $\Gamma_\nu$ is called
the active neutrino damping rate. 
As \eq\nr{master_simple} shows, $\im\Sigma^0 > 0$ 
also leads to a positive sterile neutrino production rate.

%
\subsection{Relation of active neutrino and mesonic spectral functions}
\la{subse:relation}

With the conventions set, we need to 
determine $i \bsl{\tilde\Sigma}_{\alpha\alpha}(\tilde Q)$. 
A simple computation to second order in the Fermi interaction 
(as already mentioned, the first-order contribution vanishes
in the absence of chemical potentials) yields
\be
 i \bsl{\tilde\Sigma}_{\alpha\alpha}(\tilde Q) = 
 4 G_F^2 \sum_{H = W,Z}
 \Tint{\tilde R_\bo}
 p_H \, \tilde \gamma_\mu \frac{i \bsl{\tilde Q} + i \bsl{\tilde R}}
 {(\tilde Q + \tilde R)^2 + m_{l_H}^2} 
 \tilde \gamma_\nu\, \tilde C^H_{\mu\nu}(\tilde R)
 \;, \la{raw1}
\ee
where $p_W \equiv 2$, $p_Z \equiv 1/2$ are the ``weights''
of the charged and neutral current channels; 
$m_{l_W} \equiv m_{l_\alpha}$ is the mass of the 
charged lepton of generation $\alpha$; 
$m_{l_Z} \equiv m_{\nu_\alpha} = 0$ is the mass of the 
MSM active neutrino;  
$\tilde R_\bo \equiv(\rr,\vec{r})$ denotes bosonic Matsubara four-momenta;
and we have defined the Euclidean charged and neutral current
correlators in accordance with \eq\nr{bE}, {\it viz.}
\be
 \tilde C_{\mu\nu}^W(\tilde R) \equiv 
 \int_{\tilde x}
  e^{i \tilde R\cdot (\tilde x - \tilde y)}
 \Bigl\langle
  \tilde H^W_\mu(\tilde x) \tilde H_\nu^{W^\dagger} (\tilde y)
 \Bigr\rangle 
 \;, \quad
 \tilde C_{\mu\nu}^Z(\tilde R) \equiv 
 \int_{\tilde x} e^{i \tilde R\cdot (\tilde x - \tilde y)}
 \Bigl\langle
  \tilde H^Z_\mu(\tilde x) \tilde H_\nu^{Z} (\tilde y)
 \Bigr\rangle 
 \;, \la{CW}
\ee
where $\tilde x_\mu \equiv (\tilde x^0, x^i)$,
$\tilde r_\mu \equiv(\rr,\tilde r_i) \equiv (\rr,-r^i)$, 
$\tilde x\cdot \tilde R \equiv \tilde x_\mu\tilde r_\mu =
\tilde x^0 \tilde r_0 - x^i r^i$, and 
$\int_{\tilde x} \equiv \int_0^\beta \! {\rm d} \tilde x^0  
 \int \! {\rm d}^3\vec{x}$. 
The tildes in $\tilde C$'s and $\tilde H$'s 
remind us of the fact that we are for 
the moment using Euclidean Dirac-matrices in the hadronic currents.
We also point out that the Dirac algebra remaining in \eq\nr{raw1}
cannot in general be greatly simplified, since the functions
$\tilde C^H_{\mu\nu}(\tilde R)$ can in principle contain both 
symmetric (e.g.~$\tilde R_\mu \tilde R_\nu$) and antisymmetric 
($\epsilon_{\alpha\beta\mu\nu} \tilde u_\alpha \tilde R_\beta$) structures
in $\mu\leftrightarrow\nu$.

It is important to stress now that in \eq\nr{raw1} we have assumed
a free form for the lepton/neutrino propagators inside the loop. 
Naturally, one could also allow for a general structure, such as 
the one in \eq\nr{prop}, to appear here. To keep the discussion
as simple as possible, however, we treat the inner lines to 
zeroth order in $G_F$ in this paper.

To see how \eq\nr{raw1} can be analysed, let us simplify 
the notation somewhat. The essential issue is what happens
with the Matsubara frequencies, and we hence rewrite the 
structure as 
\be
 i \bsl{\tilde\Sigma} (\qq,\vec{q}) \equiv 
 \Tint{\tilde R_\bo} 
 \frac{\tilde f_{\mu\nu}( i [\qq + \rr] ) }
 {(\qq + \rr)^2 + E_1^2}
 \;\tilde C^H_{\mu\nu}(\rr,\vec{r})
 \qquad\qquad
 \TopoSBnew(\Lgh,\Bqu,\Aqu)
 \;, \la{GE}
\ee
where 
\be
 E_1 \equiv \sqrt{(\vec{q} + \vec{r})^2 + m_{l_H}^2} 
 \;. \la{shorts}
\ee
The drawing in \eq\nr{GE} illustrates the momentum flow, 
with the thick line indicating the composite mesonic propagator.
The challenge is simply to rewrite \eq\nr{GE} in a form where 
the analytic continuation needed for $\re\Sigma$ and $\im\Sigma$
can be carried out in a controlled way, 
without generating any non-converging sums or integrals.

In order to proceed, we first rewrite \eq\nr{GE} as 
\be
 i \bsl{\tilde\Sigma} (\qq,\vec{q})
 = \int_{\vec{r}} T \sum_{\rrbo} \sum_{\ssfe}
 \delta_{\tss-\qq-\rr} 
 \frac{\tilde f_{\mu\nu}( i \tss )}{\tss^2 + E_1^2}
 \tilde C^H_{\mu\nu}(\rr,\vec{r})
 \;, 
\ee
where $\int_{\vec{r}} \equiv \int\! {\rm d}^3 \vec{r} / (2\pi)^3$, 
and $\rrbo,\ssfe$ denote bosonic and fermionic Matsubara
frequencies, respectively. 
Furthermore, we note that the Kronecker $\delta$-function can be 
expressed as
\be
 \delta_{\tss-\qq-\rr} = T \int_0^\beta\! {\rm d}\tau \, 
 e^{i \tau (\tss-\qq-\rr)}
 \;. 
\ee
Thereby the correlation function becomes 
\be
 i \bsl{\tilde \Sigma} (\qq,\vec{q})
 = 
 \int_{\vec{r}} \int_0^\beta \! {\rm d}\tau \, e^{-i \tau \qq}
 \biggl[
  T \sum_{\ssfe} \frac{\tilde f_{\mu\nu} ( i \tss )}
                        {\tss^2 + E_1^2} e^{i \tau \tss} 
 \biggr]
 \biggl[ 
  T \sum_{\rrbo} e^{-i \tau \rr}  \tilde C^H_{\mu\nu}(\rr,\vec{r})
 \biggr] 
 \;. \la{factorised}
\ee
The sum inside the first square brackets 
can be performed according to \eq\nr{fsum}. 
In order to handle the second square brackets, 
we express $\tilde C^H_{\mu\nu}(\rr,\vec{r})$ through 
the spectral representation in \eq\nr{bErhorel}.
Inserting this into \eq\nr{factorised} and changing
orders of integration, we obtain
\ba
 i \bsl{\tilde \Sigma} (\qq,\vec{q}) & = & 
 \int_{\vec{r}}
 \frac{\nF{}(E_1)}{2 E_1}
 \int_{-\infty}^{\infty}\frac{{\rm d} \omega}{\pi}
 \tilde \rho^H_{\mu\nu}(\omega,\vec{r})
 \times \nn 
 & \times &
 \int_0^\beta \! {\rm d}\tau \, e^{-i \tau \qq}
 \Bigl[
   \tilde f_{\mu\nu}(-E_1) e^{(\beta - \tau) E_1} 
-  \tilde f_{\mu\nu}( E_1) e^{\tau E_1} 
 \Bigr]
  T \sum_{\rrbo} \frac{e^{-i \tau \rr}}{\omega - i \rr} 
 \;. \hspace*{0.8cm} \la{intermediate}
\ea
As the next step, the sum in \eq\nr{intermediate} 
can be performed. In fact, 
the result can immediately be obtained by choosing 
suitable constants $c,d$ in \eq\nr{bsum}:
for $0 < \tau < \beta$, 
\be
 T \sum_{\rrbo} \frac{e^{-i \tau \rr}}{\omega - i \rr} = 
 \nB(\omega) e^{(\beta - \tau) \omega} 
 \;,
\ee
where $\nB(x)\equiv 1/[\exp(\beta x) - 1]$.
The integral over $\tau$ can then be carried out
according to \eq\nr{fint}, leading finally to 
\ba
 i \bsl{\tilde \Sigma} (\qq,\vec{q}) & = & 
 \int_{\vec{r}}
 \frac{\nF{}(E_1)}{2 E_1}
 \int_{-\infty}^{\infty}\frac{{\rm d} \omega}{\pi}
 \tilde \rho^H_{\mu\nu}(\omega,\vec{r}) \nB(\omega)
 \times \nn 
 & \times &
 \biggl[
   \tilde f_{\mu\nu}(-E_1)  
   \frac{ e^{\beta(\omega + E_1)} + 1}{ i \qq + \omega + E_1} 
-  \tilde f_{\mu\nu}( E_1) 
   \frac{ e^{\beta \omega} + e^{\beta E_1}}{ i \qq + \omega - E_1} 
 \biggr]
 \;. \hspace*{0.8cm} \la{fullSigma}
\ea

It remains to carry out the analytic continuation 
$\qq \to -i [q^0 \pm i 0^+]$ and to take the real and imaginary parts. 
In particular, taking the imaginary part as defined by \eq\nr{Discdef} 
goes with \eq\nr{delta}, given that $\qq$ only appears 
in a simple way in \eq\nr{fullSigma}.
Carrying then out the integration over $\omega$ to remove the 
$\delta$-functions, and returning simultaneously to Minkowskian
Dirac-matrices 
($
 \tilde \gamma_\mu \tilde \gamma_\mu = \gamma^\mu \gamma_\mu
$), 
whereby the tildes can be removed, we arrive at 
\ba
 \im\bsl{\Sigma}(q^0,\vec{q}) & = & \int_{\vec{r}} 
 \frac{\cosh({\beta q^0}/{2})}{4 E_1 \cosh({\beta E_1}/{2})} 
 \times \nn & \times & 
 \biggl\{
  \frac{f^{\mu\nu}(-E_1) \rho^H_{\mu\nu}(-q^0 - E_1,\vec{r})}
  {\sinh[{\beta(q^0 + E_1)}/{2}]}
  - 
  \frac{f^{\mu\nu} (E_1) \rho^H_{\mu\nu}(-q^0 + E_1,\vec{r}) }
 {\sinh[{\beta(q^0 - E_1)}/{2}]}
 \biggr\}
 \;. \hspace*{0.5cm} \la{final}
\ea
Defining 
\be
 \Delta(p^0,\vec{p},m) \equiv \bsl{P} + m 
 \;,  
\ee
and reintroducing masses in the numerators to remind us of the fact
that the $\Delta$-functions are to be evaluated at the on-shell points
(the masses are in any case deleted by the chiral projectors), 
we can return to the complete notation: 
\ba
 \im\bsl{\Sigma}_{\!\alpha\alpha}(q^0,\vec{q}) & = & 4 G_F^2 \sum_{H=W,Z}
 p_H\, 
 \int \! \frac{{\rm d}^3 \vec{r}}{(2\pi)^3} \,
 \frac{\cosh({\beta q^0}/{2})}{4 E_1 \cosh({\beta E_1}/{2})}
 \times \nn  & \times & \!
 \biggl[ 
 \frac{\gamma^\mu \Delta(-E_1,\vec{q}+\vec{r},-m_{l_H})\gamma^\nu}
 {\sinh[{\beta(q^0 + E_1)}/{2}]} 
 \rho^H_{\mu\nu}(-q^0-E_1,\vec{r})
 - (E_1 \to -E_1)
 \biggr]
 \;. \hspace*{0.8cm} \la{theend} 
\ea
Here $E_1$ is from \eq\nr{shorts}. 
Let us note that $\rho^H_{\mu\nu}$ vanishes at zero frequency, so that 
the poles originating from 
the $\sinh$-functions in \eq\nr{theend} are harmless.

To conclude, $\im\Sigma$ can be 
expressed as a three-dimensional 
spatial integral, with certain hyperbolic weights, 
over the mesonic spectral functions $\rho^H_{\mu\nu}$ 
related to the charged and neutral currents.  
On the other hand, $\re\Sigma$ requires a four-dimensional
principal value integration over the same spectral functions, 
as dictated by \eqs\nr{fullSigma}, \nr{delta}.

%
\subsection{Reduction of mesonic spectral functions}

If we make certain assumptions about the quark mass matrix, 
the mesonic correlators in \eq\nr{CW}, as well as the corresponding
spectral functions $\rho^H_{\mu\nu}$ that appear
in \eq\nr{theend}, can be reduced to a small set of
quantities that have been widely addressed in the literature.

To start with, it is probably a good approximation
at temperatures 100~MeV $\lsim T \lsim$ 400 MeV to treat
the three lightest quarks as degenerate, with a certain 
mass $m_\rmi{q}$, while the three heaviest quarks can be 
assumed infinitely heavy, and ignored. In this limit the 
theory has an exact SU$_V$(3) flavour symmetry, which guarantees that 
we can split the correlators into flavour singlets and non-singlets. 
Defining $T^a$ to be traceless and Hermitean, 
and $T^0$ to be the $3\times 3$ unit matrix, we can 
then construct the flavour non-singlet and singlet 
vector and axial currents, 
\be
 \tilde V^a_\mu \equiv \bar\psi \tilde \gamma_\mu T^a \psi
 \;, \quad 
 \tilde V^0_\mu \equiv \bar\psi \tilde \gamma_\mu T^0 \psi
 \;, \quad 
 \tilde A^a_\mu \equiv \bar\psi \tilde \gamma_\mu \gamma_5 T^a \psi
 \;, \quad 
 \tilde A^0_\mu \equiv \bar\psi \tilde \gamma_\mu \gamma_5 T^0 \psi
 \;. 
\ee   
In general, the flavour symmetry allows for six 
correlation functions: 
\ba
 \tr[T^a T^b] \tilde C^V_{\mu\nu}(\tilde R) & \equiv & 
 \int_{\tilde x} e^{i \tilde R\cdot \tilde x}
 \Bigl\langle \tilde V^a_\mu(\tilde x) \tilde V^b_\nu(0) 
 \Bigr\rangle 
 \;, \\ 
 \tr[T^a T^b] \tilde C^A_{\mu\nu}(\tilde R) & \equiv & 
 \int_{\tilde x} e^{i \tilde R\cdot \tilde x}
 \Bigl\langle \tilde A^a_\mu(\tilde x) \tilde A^b_\nu(0) 
 \Bigr\rangle 
 \;, \\ 
 \tr[T^a T^b] \tilde C^{V\!\!A}_{\mu\nu}(\tilde R) & \equiv & 
 \int_{\tilde x} e^{i \tilde R\cdot \tilde x}
 \Bigl\langle \tilde V^a_\mu(\tilde x) \tilde A^b_\nu(0) 
 + \tilde A^a_\mu(\tilde x) \tilde V^b_\nu(0)
 \Bigr\rangle 
 \;, \\ 
 \tilde C^{V0}_{\mu\nu}(\tilde R) & \equiv & 
 \int_{\tilde x} e^{i \tilde R\cdot \tilde x}
 \Bigl\langle \tilde V^0_\mu(\tilde x) \tilde V^0_\nu(0) 
 \Bigr\rangle 
 \;, \\ 
 \tilde C^{A0}_{\mu\nu}(\tilde R) & \equiv & 
 \int_{\tilde x} e^{i \tilde R\cdot \tilde x}
 \Bigl\langle \tilde A^0_\mu(\tilde x) \tilde A^0_\nu(0) 
 \Bigr\rangle 
 \;, \\ 
 \tilde C^{V\!\!A0}_{\mu\nu}(\tilde R) & \equiv & 
 \int_{\tilde x} e^{i \tilde R\cdot \tilde x}
 \Bigl\langle \tilde V^0_\mu(\tilde x) \tilde A^0_\nu(0) +
 \tilde A^0_\mu(\tilde x) \tilde V^0_\nu(0)  
 \Bigr\rangle 
 \;.  
\ea
However, only four among these are non-trivial in the 
mass-degenerate limit that we are considering: the vector and 
axial currents have opposite transformation properties
in charge conjugation C, which implies that 
$\tilde C^{V\!\!A}_{\mu\nu}(\tilde R)$, 
$\tilde C^{V\!\!A0}_{\mu\nu}(\tilde R)$ vanish in this case. 

Now, the correlators that appear in \eq\nr{CW} can 
be expressed in terms of the ones just defined. We obtain 
\ba
 \tilde C^W_{\mu\nu}(\tilde R) & = & 
 \frac{|V_\rmi{ud}|^2 + |V_\rmi{us}|^2}{4}
 \Bigl[
  \tilde C^V_{\mu\nu}(\tilde R) +  
  \tilde C^A_{\mu\nu}(\tilde R) 
 \Bigr] \;, \la{CWred} \\
 \tilde C^Z_{\mu\nu}(\tilde R) & = & 
 \fr23 \Bigl[
  (1 - 2 x_\rmii{W})^2 \tilde C^V_{\mu\nu}(\tilde R) +  
  \tilde C^A_{\mu\nu}(\tilde R) 
 \Bigr] + 
 \frac{1}{36} 
  \Bigl[
  \tilde C^{V0}_{\mu\nu}(\tilde R) +  
  \tilde C^{A0}_{\mu\nu}(\tilde R) 
 \Bigr] 
 \;, \la{CZred}
\ea
where $V_\rmi{ij}$ are elements of the CKM matrix. 
Identical relations hold for the spectral functions. 

Under further assumptions, the set of independent correlators can still
be reduced. In particular, taking the chiral limit $m_\rmi{q} \to 0$,
the Ward identity following from applying an infinitesimal 
non-singlet axial transformation
on the correlator 
$\langle \tilde V^a_\mu(\tilde x) \tilde A^b_\nu(0)\rangle$  
states that $\tilde C^V_{\mu\nu}(\tilde R) = \tilde C^A_{\mu\nu}(\tilde R)$.
For the singlets this is not true in general, in spite of the 
fact that anomalous processes are in some sense 
suppressed at high temperatures (see, e.g., Ref.~\cite{nrqcd3}).
So in principle 
there remain three independent functions to determine, 
$\tilde C^{V}_{\mu\nu}$, $\tilde C^{V0}_{\mu\nu}$, $\tilde C^{A0}_{\mu\nu}$. 
 
Now, in the three-flavour theory, 
the non-singlet vector correlator $\tilde C^{V}_{\mu\nu}$
determines directly the photon spectral function~\cite{dilepton}, 
and is thus relevant for computing the photon and 
dilepton production rates, which are among
the prime observables for heavy ion collision experiments.
Therefore it has been addressed with a variety of methods in 
the literature, starting with various perturbative 
treatments~\cite{dilepton} as well as with so-called thermal 
sum rules~\cite{sums}. The (resummed~\cite{bp}) perturbative treatments
have reached a great degree of sophistication by now~\cite{htl}, with 
different strategies applicable in different parts of the phase space. 
On the other hand, it has also been realised that an analytic
continuation of imaginary time correlators defined on the finite
$\tau$-interval {\em can} be carried out in principle~\cite{acont}, 
which opens up the possibility of lattice QCD determinations. 
There have indeed been attempts at practical implementations
of a certain analytic continuation from numerical data~\cite{mem}, 
though they are not without problems for the moment~\cite{mem2}. 
Finally, for $T\ll 150$~MeV, chiral perturbation theory can be 
systematically applied~\cite{xpt} or at least used as a solid baseline
for the computation of $\tilde C^{V}_{\mu\nu}$
(and thus of $\rho^V_{\mu\nu}$)~\cite{ubw}. 

The vector singlet $\tilde C^{V0}_{\mu\nu}$, 
on the other hand, can be associated with 
the baryon number current, whose susceptibility $\chi$ (which is 
an integral over the spectral function, 
$
 \chi = \int_{-\infty}^{\infty} \! {\rm d} \omega \,
 \rho^{V0}_{00} (\omega,\vec{0}) / \pi \omega
$) is argued
to be relevant for so-called event-to-event fluctutations
in heavy ion collision experiments~\cite{hisuscs}.
In resummed perturbative treatments~\cite{av} 
the difference between the vector singlet and non-singlet 
susceptibilites (this difference is often 
called the ``off-diagonal quark number susceptibility'') 
is however very small, 
being suppressed by $\alpha_s^3\ln(1/\alpha_s)$~\cite{bir}, 
so that at high enough temperatures $\tilde C^{V0}_{\mu\nu}$ can well be 
approximated (up to an overall factor) by $\tilde C^{V}_{\mu\nu}$.
At lower temperatures close to $T\simeq 150$~MeV, on the other hand, 
a lattice determination would again be needed; unfortunately, 
the difference between $\tilde C^{V0}_{\mu\nu}$ and
$\tilde C^{V}_{\mu\nu}$ is given by a disconnected quark-line 
contraction which is technically rather difficult to measure 
accurately~\cite{suscs}. We should of course stress that the 
susceptibility alone contains much less information than the 
complete function $\tilde C^{V0}_{\mu\nu}$, or its analytic
continuation $\rho^{V0}_{\mu\nu}$. The general high-temperature
structure of the latter has been analysed in Ref.~\cite{ay}. 
Finally, at very low temperatures,
chiral perturbation theory predicts that 
$\tilde C^{V0}_{\mu\nu}$ is strongly suppressed with
respect to $\tilde C^{V}_{\mu\nu}$.

Much the same comments can be made for 
the axial singlet current, $\tilde C^{A0}_{\mu\nu}$.
At high enough temperatures, where resummed QCD perturbation
theory is applicable, it agrees up to a certain order in 
the resummed perturbative expansion with $\tilde C^{V0}_{\mu\nu}$. 
At lower temperatures, the difference between $\tilde C^{A0}_{\mu\nu}$
and $\tilde C^{V0}_{\mu\nu}$ becomes significant. This
difference is of course quite interesting in its own right, being
related to the $\eta'$-meson and the chiral anomaly. However, 
analytic and numerical treatments are demanding in this range.
At very low temperatures, chiral perturbation theory 
indicates that $\tilde C^{A0}_{\mu\nu}$ remains more significant 
than $\tilde C^{V0}_{\mu\nu}$.

%
\subsection{Perturbative limit}

Returning finally to the simplest possible logic, we wish
to inspect the mesonic spectral function $\rho^H_{\mu\nu}$ 
in naive perturbation theory. The purpose is to show that 
in this case, \eq\nr{theend} leads to the familiar structure
of Boltzmann equations. 
(The procedure we follow is analogous to the one first 
worked out for simpler cases in Ref.~\cite{weldon2}, 
and also used in previous active neutrino damping rate 
computations~\cite{damping}.)

In order to reach this goal, we write  generic hadronic 
currents (\eqs\nr{HW}, \nr{HZ}) in the form $\tilde H_\mu(\tilde x) = 
\bar q_2(\tilde x) \tilde \gamma_\mu \Gamma q_3(\tilde x)$, where
$q_2, q_3$ denote quark fields and 
$\Gamma$ is some Dirac matrix structure.  
Carrying then out the contractions
in the correlators of \eq\nr{CW}, we obtain 
\be
 \tilde C^H_{\mu\nu}(\tilde R) = 
 - \Nc\, \Tint{\tilde T_\fe}
 \tr\biggl\{
   \frac{-i(\bsl{\tilde T} + \bsl{\tilde R}) + m_2}
   {(\tt + \rr)^2 + E_2^2}
   \, \tilde \gamma_\mu \Gamma \,
   \frac{-i\bsl{\tilde T} + m_3}
   {\tt^2 + E_3^2}
   \, \tilde \gamma_\nu \Gamma
 \biggr\}
 \;,  \la{tCHex}
\ee
where $\Nc = 3$ is the number of colours, and
\be
 E_2 \equiv \sqrt{(\vec{t}+\vec{r})^2 + m_2^2} 
\;, \quad
 E_3 \equiv \sqrt{\vec{t}^2 + m_3^2}
 \;.
 \la{E2E3}
\ee
The important issue is again what happens with the
Matsubara frequencies. Following the steps
in Section~\ref{subse:relation}, 
and denoting the complete numerator of \eq\nr{tCHex}
by a function $\tilde g_{\mu\nu}$, 
we obtain
\ba
 &   &  
 \hspace*{-2cm}
 T\sum_{\ttfe} \frac{\tilde g_{\mu\nu}(i\tt + i \rr,i\tt)}
 {[(\tt + \rr)^2 + E_2^2][\tt^2 + E_3^2]} \nn 
 & = & 
 T \sum_{\ttfe,\uufe} \delta_{\uu-\tt-\rr} 
 \frac{\tilde g_{\mu\nu}(i\uu,i\tt)}{[\uu^2 + E_2^2][\tt^2 + E_3^2]}
 \nn 
 & = & 
 \int_0^\beta\! {\rm d}\tau \, e^{-i\tau\rr}
 \; T\sum_{\uufe} e^{i\uu\tau}
 \; T\sum_{\ttfe} e^{-i\tt\tau} 
 \frac{\tilde g_{\mu\nu}(i\uu,i\tt)}{[\uu^2 + E_2^2][\tt^2 + E_3^2]}
 \nn
 & = & 
 \frac{\nF{}(E_2)\nF{}(E_3)}{4E_2E_3} 
 \int_0^\beta\! {\rm d}\tau \, e^{-i\tau\rr}
 \biggl[ 
   +\tilde g_{\mu\nu}(-E_2,+E_3) e^{(\beta-\tau)(E_2 + E_3)}
 - \nn & & 
 \hphantom{
 \int_0^\beta\! {\rm d}\tau \, e^{-i\tau\rr}
 \frac{\nF{}(E_2)\nF{}(E_3)}{4E_2E_3} } 
 - \tilde g_{\mu\nu}(-E_2,-E_3) e^{(\beta-\tau)E_2 + \tau E_3}
 - \nn &  & 
 \hphantom{
 \int_0^\beta\! {\rm d}\tau \, e^{-i\tau\rr}
 \frac{\nF{}(E_2)\nF{}(E_3)}{4E_2E_3} } 
 - \tilde g_{\mu\nu}(+E_2,+E_3) e^{(\beta-\tau)E_3 + \tau E_2}
 + \nn &  & 
 \hphantom{
 \int_0^\beta\! {\rm d}\tau \, e^{-i\tau\rr}
 \frac{\nF{}(E_2)\nF{}(E_3)}{4E_2E_3} } 
  + \tilde g_{\mu\nu}(+E_2,-E_3) e^{\tau(E_2 + E_3)}
 \biggr] 
 \;, 
\ea
where we made use of \eq\nr{fsum}. The integral remaining can be carried 
out by using \eq\nr{bint} (recalling that $\rr$ is bosonic), and the 
spectral function follows then by picking up the discontinuity
across the real axis, according to \eq\nr{delta}: 
\ba
 \tilde \rho^H_{\mu\nu}(R) & = & 
 \int_{\vec{t}} \frac{-\pi}{4 E_2 E_3} \biggl[ 
 + \delta(r^0 + E_2 + E_3) 
 \tilde g_{\mu\nu}(-E_2, +E_3) (1 - \nF{2}-\nF{3} )
 + \nn &  & 
 \hphantom{ \int_{\vec{t}} \frac{-\pi}{4 E_2 E_3} }
 + \delta(r^0 + E_2 - E_3) 
 \tilde g_{\mu\nu}(-E_2, -E_3) (\nF{2}-\nF{3} )
 + \nn &  & 
 \hphantom{ \int_{\vec{t}} \frac{-\pi}{4 E_2 E_3} }
 + \delta(r^0 - E_2 + E_3) 
 \tilde g_{\mu\nu}(+E_2, +E_3) (\nF{3}-\nF{2} )
 + \nn &  & 
 \hphantom{ \int_{\vec{t}} \frac{-\pi}{4 E_2 E_3} }
 + \delta(r^0 - E_2 - E_3) 
 \tilde g_{\mu\nu}(+E_2, -E_3) (\nF{2}+\nF{3} -1 )
 \biggr] \;,  \la{rhoH}
\ea 
where we have denoted $\nF{i} \equiv \nF{}(E_i)$.

The expression in \eq\nr{rhoH} can now be inserted 
into \eq\nr{theend}. There are eight different terms in total. 
The hyberbolic functions in \eq\nr{theend} can be 
rewritten in terms of $\nF{}$'s and $\nB$'s, and making
use of relations such as 
\be
 \delta(-q^0 + E_1 + E_2 + E_3)\nB(q^0 - E_1)(1-\nF{2} - \nF{3})
 = 
 \delta(-q^0 + E_1 + E_2 + E_3) \nF{2} \nF{3}
 \;, 
\ee
they can be reorganised in a simpler form. In view of \eq\nr{master},
we also choose to factor out $\nF{}^{-1}(q^0)$ from all the terms. 
After a number of trivial if tedious manipulations, we
finally arrive at 
\ba
 & &  \hspace*{-1cm}
 \im\bsl{\Sigma}(Q) = 2 \Nc G_F^2 \nF{}^{-1}(q^0)  \sum_{H = W,Z} p_H 
 \int \! \frac{{\rm d}^3 \vec{p}_1}{(2\pi)^3 2 E_1} \,
 \int \! \frac{{\rm d}^3 \vec{p}_2}{(2\pi)^3 2 E_2} \,
 \int \! \frac{{\rm d}^3 \vec{p}_3}{(2\pi)^3 2 E_3} \,
 \times  
 \nn 
 & \times \biggl\{ & \!\!\!
 (2\pi)^4 \delta^{(4)}(P_1+P_2+P_3-Q )\, 
 \nF{1}\nF{2}\nF{3} \,
 \mathcal{A}(-m_{l_H},m_2,-m_3)
 + \ScatA \nn & + & \!\!\!
 (2\pi)^4 \delta^{(4)}(P_2+P_3 -P_1-Q) \,
 \nF{2}\nF{3}(1-\nF{1}) \,
 \mathcal{A}(m_{l_H},m_2,-m_3)
 + \ScatB \nn & + & \!\!\!
 (2\pi)^4 \delta^{(4)}(P_1+P_3-P_2-Q) \, 
 \nF{1}\nF{3}(1-\nF{2}) \,
 \mathcal{A}(-m_{l_H},-m_2,-m_3)
 + \ScatC \nn & + & \!\!\!
 (2\pi)^4 \delta^{(4)}(P_1+P_2-P_3-Q) \, 
 \nF{1}\nF{2}(1- \nF{3}) \,
 \mathcal{A}(-m_{l_H},m_2,m_3)
 + \ScatD \nn & + & \!\!\!
 (2\pi)^4 \delta^{(4)}(P_1-P_2-P_3-Q) \,
 \nF{1}(1-\nF{2})(1-\nF{3}) \,
 \mathcal{A}(-m_{l_H},-m_2,m_3)
 + \ScatE \nn & + & \!\!\!
 (2\pi)^4 \delta^{(4)}(P_2-P_1-P_3-Q) \, 
 \nF{2}(1-\nF{1})(1-\nF{3}) \,
 \mathcal{A}(m_{l_H},m_2,m_3)
 + \ScatF \nn & + & \!\!\!
 (2\pi)^4 \delta^{(4)}(P_3 -P_1-P_2-Q) \,
 \nF{3} (1-\nF{1})(1-\nF{2}) \,
 \mathcal{A}(m_{l_H},-m_2,-m_3)
 + \ScatG \nn & + & \!\!\!
 (2\pi)^4 \delta^{(4)}(-P_1-P_2-P_3-Q ) \, 
 (1-\nF{1})(1-\nF{2})(1-\nF{3}) \,
 \mathcal{A}(m_{l_H},-m_2,m_3)
 \biggr\}
 , \ScatH  \nn 
 \la{pertB}
\ea
where 
\be
 \mathcal{A}(m_{l_H},m_2,m_3) \equiv
 \gamma^{\mu} (\bsl{P}_{\! 1} + m_{l_H}) \gamma^\nu
 \, \tr\Bigl[
  (\bsl{P}_{\! 2} + m_2) \gamma_\mu \Gamma 
  (\bsl{P}_{\! 3} + m_3) \gamma_\nu \Gamma 
 \Bigr]
 \;.
\ee
Here $P_i \equiv (E_i,\vec{p}_i)$ are on-shell four-momenta, 
and the energies have been transformed from \eqs\nr{shorts}, \nr{E2E3} into
\be
 E_1 \equiv 
 \sqrt{\vec{p}_1^2 + m_{l_H}^2}
 \;, \quad
 E_2 \equiv \sqrt{\vec{p}_2^2 + m_2^2}
 \;, \quad
 E_3 \equiv \sqrt{\vec{p}_3^2 + m_3^2}
 \;. \la{Es}
\ee
The graphs in \eq\nr{pertB} illustrate the various processes, with 
time and momenta assumed to run from left to right, and arrows
indicating particles / antiparticles in the usual way. 
The general structure is clearly
what we expect from Boltzmann equations, however we have arrived
at it without any model assumptions, evaluations of scattering
matrix elements, or spin averages. 
It is easy to check 
(by making use of the properties of the $\nF{}$'s as well as the 
fact that the first argument of $\mathcal{A}(m_1,m_2,m_3)$ can 
be dropped as it will in any case be projected out by $a_L, a_R$, 
and that $\mathcal{A}(0,-m_2,-m_3) = \mathcal{A}(0,m_2,m_3)$  
for $\Gamma = a + b \gamma_5$)
that the expression in \eq\nr{pertB} 
is symmetric in $Q\to -Q$, as it should be.

For illustration 
it is useful to take one more step, and insert this result
into the simplified form in \eq\nr{master_simple}. 
Let us consider the contribution from the third term in \eq\nr{pertB}, 
for instance, and choose the charged current, i.e.\ $H\equiv W$. 
Then $\Gamma = a_L$, and we obtain 
\ba
 & &  \hspace*{-1.8cm}
   \frac{{\rm d} N_I(x,\vec{q})}{{\rm d}^4 x\,{\rm d}^3 \vec{q}}
 =  \frac{16 \Nc G_F^2}{(2\pi)^3 2 q^0}
   \sum_{\alpha = 1}^{3}  
   \frac{|M_D|^2_{\alpha I}}{M_I^2}
  \int \! \frac{{\rm d}^3 \vec{p}_1}{(2\pi)^3 2 E_1} \,
  \int \! \frac{{\rm d}^3 \vec{p}_2}{(2\pi)^3 2 E_2} \,
  \int \! \frac{{\rm d}^3 \vec{p}_3}{(2\pi)^3 2 E_3} \,
  \times  
  \nn 
  & \times & \!\!\!
  (2\pi)^4 \delta^{(4)}(P_1+P_3-P_2-Q) \,
  \nF{}(u\cdot P_1)\nF{}(u \cdot P_3)[1-\nF{}(u \cdot P_2)] 
  \times   \nn 
 & \times & \!\!\!
 \tr\Bigl[ \bsl{Q} a_L 
  \gamma^{\mu} (\bsl{P}_{\! 1} - m_{\nu_\alpha}) \gamma^\nu a_R
  \Bigr]
 \tr\Bigl[
  (\bsl{P}_{\! 2} - m_2) \gamma_\mu a_L
  (\bsl{P}_{\! 3} - m_3) \gamma_\nu a_L
 \Bigr] \;, \ScatC
 \la{example}
\ea
where $m_{\nu_\alpha} = 0$ has only been kept to formally
indicate the anti-particle direction of the line, and we 
have written everything in a Lorentz-covariant form. 
All explicit masses drop out because of the chiral projectors. 
The Dirac algebra is elementary, 
\be
  \tr\Bigl[ \bsl{Q}  
  \gamma^{\mu} \bsl{P}_{\! 1} \gamma^\nu a_R
  \Bigr]
 \tr\Bigl[
  \bsl{P}_{\! 2}  \gamma_\mu 
  \bsl{P}_{\! 3}  \gamma_\nu a_L
 \Bigr] = 16\, Q \cdot P_3 \, P_1 \cdot P_2
 \;.  \la{ex2}
\ee
This expression is positive, and carrying out the phase-space 
integral in \eq\nr{example}, one obtains a finite function of $|\vec{q}|$. 
Let us remark, though, that while the integration over the energy-conservation
constraint is simple in the center-of-mass frame, 
the plasma four-velocity $u$ becomes non-trivial if this 
frame is chosen; in general, therefore, the phase-space integrals 
that appear in \eq\nr{example} are technically non-trivial. 

It is appropriate to end now by pointing out  that \eqs\nr{example},
\nr{ex2} contain significant differences with respect to the
spin-averaged matrix elements squared that appear in active neutrino
scattering cross-sections~\cite{damping}, and have sometimes also
been inserted into the Boltzmann equations for sterile neutrino
production. In particular, the left-most trace in \eq\nr{ex2}
contains the projector
$a_R$,  rather than $a_L$; this is because \eq\nr{master} contains
the active neutrino propagator, rather than the self-energy. As a
consequence, \eq\nr{ex2} does not lead to a purely $s$-channel
momentum structure  ($16\, Q \cdot P_2 \, P_1 \cdot P_3$) like the
active neutrino scattering cross sections~\cite{damping}. Whether
this special example has any practical significance is not clear at
this stage, but it  illustrates  the advantages of our framework
where the correct structures are  produced automatically from thermal
field theory, rather than  having to be inserted by hand.

%
\section{Conclusions and Outlook}
\la{se:concl}

While the sterile neutrino production rate has previously been 
investigated in the literature in some detail, the hadronic contributions 
to it have never been analysed properly. These contributions involve strongly
interacting dynamics at temperatures of the order of the QCD scale, where 
neither perturbation theory nor the dilute hadronic gas approximation 
are valid. 

To confront this situation, we have derived 
a general relation that expresses the sterile neutrino
production rate in terms of the active neutrino spectral function,
computable by using equilibrium thermal field theory within the
Minimal Standard  Model (\eq\nr{master}). The active neutrino
spectral function can in turn be expressed in terms of the real and 
imaginary parts of the active neutrino self-energy (\eq\nr{master2}).
These equations show that hadronic contributions may play a role 
in three different ways: 
\begin{itemize}

\item[(1)] 
Most importantly, the hadronic degrees of freedom contribute
at leading order to the imaginary part of the active neutrino
self-energy, $\im\Sigma$. 
Therefore, we have expressed the hadronic contribution
to $\im\Sigma$ as a certain convolution of the spectral functions 
related to hadronic current-current correlation functions (\eq\nr{theend}). 
The latter can in turn be expressed  in terms of standard vector and 
axial-vector correlators  (\eqs\nr{CWred}, \nr{CZred}) that can be studied 
with a number of different theoretical methods, and are also partly related 
to experimental observables addressed in the heavy ion program.  

\item[(2)]
The parametrically dominant contribution to the real part of the
active neutrino self-energy, $\re\Sigma$,  arises from 1-loop graphs 
within the electroweak theory, and does not contain hadronic effects. 
On the other hand, subdominant contributions, formally suppressed 
by $\alpha_w$, do contain hadronic effects. 
As shown by \eq\nr{fullSigma}, these contributions can be  
expressed in terms of a certain weighted integral over the same hadronic 
spectral functions that appear in the imaginary part.  

\item[(3)]
Finally, the sterile neutrino production rate equation contains
a time derivative ${\rm d}/{\rm d}t$, the Hubble rate $H$, 
and the temperature $T$ (cf.~\eq\nr{expansion}). 
In cosmology, all of these are related via the Einstein equations. 
The relation is again sensitive to hadronic effects, via the 
equation-of-state of the primordial plasma. The current status
and phenomenological fits for all the thermodynamic functions
that appear in the time-temperature relation can be found
in Sec.~IV of Ref.~\cite{phen}.

\end{itemize}

To summarise, our formulae should allow to estimate for the first time
the systematic hadronic uncertainties that exist in computations of 
sterile neutrino production through active-sterile transitions. 
A numerical evaluation of these effects is in progress.

Apart from these non-perturbative aspects, we have also  demonstrated
that our general formulae allow to derive, without further
assumptions, the appropriate Boltzmann equations that  apply in the
naive weak-coupling limit (\eq\nr{pertB}). It may  in fact be useful
to repeat our computations for the  leptonic contributions as well,
since a first-principles derivation  frees us from the need to argue
about spin averages or symmetry  factors, and produces automatically
the  correct Dirac and chiral structures for the Boltzmann equations. 

%
\section*{Acknowledgements}

The work of T.A.\ was supported in part by the grants-in-aid from 
the Ministry of Education, Science, Sports, and Culture of Japan,
Nos.\ 16081202 and 17340062, 
and that of M.S.\ by the Swiss National Science Foundation.


\appendix
\renewcommand{\thesection}{Appendix~\Alph{section}}
\renewcommand{\thesubsection}{\Alph{section}.\arabic{subsection}}
\renewcommand{\theequation}{\Alph{section}.\arabic{equation}}

%
\section{Basic relations for bosons}

We list in this Appendix some common definitions and relations 
that apply to two-point correlation functions built out of bosonic 
operators; for more details see, e.g., Refs.~\cite{old,mlb}. 

We denote Minkowskian space-time coordinates by $x=(t,x^i)$ and
momenta by $Q = (q^0,q^i)$, while their Euclidean counterparts
are denoted by $\tilde x = (\tau,x^i)$, 
$\tilde Q = (\qq,q_i)$. Wick rotation is carried 
out by $\tau \leftrightarrow i t$, 
$\qq \leftrightarrow - i q^0$. Arguments of operators
denote implicitely whether we are in Minkowskian or Euclidean space-time.
In particular, Heisenberg-operators are defined as
\be
 \hat O(t,\vec{x}) \equiv 
 e^{i \hat H t} \hat O(0,\vec{x}) e^{- i \hat H t}
 \;, \quad
 \hat O(\tau,\vec{x}) \equiv 
 e^{\hat H \tau} \hat O(0,\vec{x}) e^{- \hat H \tau}
 \;.
\ee
The thermal ensemble is defined by the density matrix
$\hat\rho = Z^{-1} \exp(-\beta \hat H)$.

\renewcommand{\A}{\hat\phi_\alpha}
\renewcommand{\B}{\hat{\phi}^\dagger_\beta}
\newcommand{\I}{\int\!{\rm d}t\,{\rm d}^3 \vec{x}\,e^{i Q\cdot x}}

We denote the operators which appear in the two-point functions
by $\A(x)$, $\B(x)$. They could
be elementary field operators,
but they could also be composite operators consisting of
a product of elementary field operators. 

We can now define various classes of 
correlation functions. The ``physical'' correlators are defined as
\ba
 \Pi^{>}_{\alpha\beta}(Q) & \equiv & 
 \I \Bigl\langle \A(x) \B(0) \Bigl\rangle
 \;,   
 \la{bL}
 \\
 \Pi^{<}_{\alpha\beta}(Q) & \equiv & 
 \I \Bigl\langle \B(0)  \A(x) \Bigl\rangle
 \;,   
 \\
 \la{bS}
 \rho_{\alpha\beta}(Q) & \equiv & 
 \I \Bigl\langle \fr12 \Bigl[ \A(x) , \B(0) \Bigr] \Bigl\rangle
 \;,   
 \la{brho}
\ea
where $\rho_{\alpha\beta}$ is called the spectral function, 
while the ``retarded''/``advanced'' correlators can be defined as 
\ba
 \Pi^R_{\alpha\beta}(Q) & \equiv & 
 i \I \Bigl\langle \Bigl[ \A(x) , \B(0) \Bigr] \theta(t) \Bigl\rangle
 \;, 
 \la{bR}
 \\
 \Pi^A_{\alpha\beta}(Q) & \equiv & 
 i \I \Bigl\langle  - \Bigl[ \A(x) , \B(0) \Bigr] \theta(-t) \Bigl\rangle
 \;. 
 \la{bA} 
\ea
On the other hand, from the computational point of view one 
is often faced with ``time-ordered'' correlation functions, 
\ba
 \Pi^T_{\alpha\beta}(Q) & \equiv & 
  \I \Bigl\langle \A(x)  \B(0) \theta(t)
                 +\B(0)  \A(x) \theta(-t) \Bigl\rangle
 \;, \la{bT}
\ea
which appear in time-dependent perturbation theory, or with 
the ``Euclidean'' correlator
\ba
 \Pi^E_{\alpha\beta}(\tilde Q) & \equiv & 
 \int_0^\beta\!{\rm d}\tau \int \! {\rm d}^3 \vec{x}\,
  e^{i \tilde Q \cdot \tilde x}
  \Bigl\langle \A(\tilde x)  \B(0) \Bigl\rangle
 \;,
 \la{bE}
\ea
which appears in non-perturbative formulations.
Note that the Euclidean correlator is also time-ordered by definition, 
and can be computed with standard imaginary-time functional integrals 
in the Matsubara formalism. 

Now, all of the correlation functions defined 
can be related to each other.
In particular, all correlators can be expressed in terms of the spectral
function, which in turn can be determined as a certain analytic 
continuation of the Euclidean correlator. In order to do this, 
we may first insert sets of energy eigenstates, to obtain
the Fourier-space version of the so-called Kubo-Martin-Schwinger (KMS) 
relation:  
$
 \Pi^{<}_{\alpha\beta}(Q) 
 = e^{-\beta q^0} \Pi^{>}_{\alpha\beta}(Q)
$.
Then 
$
 \rho_{\alpha\beta}(Q)
 = [\Pi^{>}_{\alpha\beta}(Q) - \Pi^{<}_{\alpha\beta}(Q)]/2
$ and, conversely, 
\be
 \Pi^{>}_{\alpha\beta}(Q) = 2[1 + \nB(q^0)] \rho_{\alpha\beta}(Q)
 \;, \quad
 \Pi^{<}_{\alpha\beta}(Q) = 2 \nB(q^0) \rho_{\alpha\beta}(Q)
 \;, \la{bLSrel}
\ee
where $\nB(x)\equiv 1/[\exp(\beta x) - 1]$. Inserting the representation
\be
 \theta(t) = i \int_{-\infty}^{\infty} \! \frac{{\rm d}\omega}{2\pi}
 \frac{e^{-i\omega t}}{\omega + i 0^+}
 \la{theta}
\ee
into the definitions of $\Pi^R$, $\Pi^A$, we obtain
\be
 \Pi^R_{\alpha\beta}(Q) 
 =  \int_{-\infty}^{\infty} \! \frac{{\rm d}\omega}{\pi} 
 \frac{\rho_{\alpha\beta}(\omega,\vec{q})}{\omega -q^0- i 0^+}
 \;, \quad 
 \Pi^A_{\alpha\beta}(Q)
 =  \int_{-\infty}^{\infty} \! \frac{{\rm d}\omega}{\pi} 
 \frac{\rho_{\alpha\beta}(\omega,\vec{q})}{\omega -q^0+ i 0^+}
 \;. 
\ee
Doing the same with $\Pi^T$ and making use of 
\be
 \frac{1}{\Delta \pm i0^+} = P\Bigl(\frac{1}{\Delta}\Bigr) 
 \mp i \pi \delta(\Delta)
 \;,  \la{delta}
\ee
produces
\be
 \Pi^T_{\alpha\beta}(Q) = 
 \int_{-\infty}^{\infty} \! \frac{{\rm d}\omega}{\pi} 
 \frac{i\rho_{\alpha\beta}(\omega,\vec{q})}{q^0 - \omega + i 0^+} + 
 2 \rho_{\alpha\beta}(q^0,\vec{q}) \nB(q^0)
 \;. 
\ee
Finally, writing the argument inside the $\tau$-integration
in \eq\nr{bE} as a Wick rotation of the integrand in \eq\nr{bL}, 
which in turn is expressed as an inverse Fourier transform of $\Pi^{>}(Q)$,
for which \eq\nr{bLSrel} is inserted, 
and changing orders of integration, we get 
\ba
 \Pi^E_{\alpha\beta}(\tilde Q) & = & 
 \int_0^\beta \! {\rm d}\tau\, e^{i \qq \tau}
 \int_{-\infty}^{\infty} \frac{{\rm d}\omega }{2\pi} e^{-\omega \tau} 
 \Pi^{>}_{\alpha\beta}(\omega,\vec{q}) =
 \int_{-\infty}^{\infty} \! \frac{{\rm d}\omega}{\pi} 
 \frac{\rho_{\alpha\beta}(\omega,\vec{q})}{\omega - i \qq}
 \;. \la{bErhorel}
\ea
This relation can formally be 
inverted by making use of \eq\nr{delta}, 
\be
 \rho_{\alpha\beta}(q^0,\vec{q}) 
 = \frac{1}{2i} \disc \Pi^E_{\alpha\beta}(\qq\to -i q^0,\vec{q})
 \;,
\ee
where the operation $\disc$ is defined in \eq\nr{Discdef}.

We also recall that bosonic Matsubara sums can be carried out through
\ba
  T \sum_{\omega_\rmi{b}} 
 \frac{i \omega_\rmi{b} c + d}
 {\omega_\rmi{b}^2 + E^2} e^{i \omega_\rmi{b} \tau}
 & \equiv &  
 (c\partial_\tau + d) 
  T \sum_{\omega_\rmi{b}} 
 \frac{e^{i \omega_\rmi{b} \tau}}
 {\omega_\rmi{b}^2 + E^2} 
 \\ 
 & = & 
 \frac{\nB(E)}{2 E} \Bigl[ 
  (-cE + d) e^{(\beta - \tau) E} + (cE + d) e ^{\tau E}
 \Bigr] \;,
 \la{bsum}
\ea
where $\omega_\rmi{b} = 2 \pi T n$,
with $n$ an integer, and we assumed $0 <  \tau < \beta$; 
and that a typical integration yields
\be
  \int_0^\beta\! {\rm d}\tau \,
 e^{-\tau (i \omega_\rmi{b} + \Delta)}
 = 
 \frac{1 - e^{-\beta\Delta}}{i \omega_\rmi{b} + \Delta} 
 \;. \la{bint}
\ee

%
\section{Basic relations for fermions}

We list in this Appendix some common definitions and relations 
that apply to two-point correlation functions built out of fermionic 
operators; for more details see, e.g., Refs.~\cite{old,mlb}.

We denote the operators which appear in the two-point functions
by $\hat\nu_\alpha(x)$, $\,\hat{\!\bar{\nu}}_\beta(x)$. They could
be elementary field operators, in which case 
the indices $\alpha,\beta$ label Dirac and/or flavour components, 
but they could also be composite operators consisting of
a product of elementary field operators. 

\renewcommand{\A}{\hat\nu_\alpha}
\renewcommand{\B}{\,\hat{\!\bar{\nu}}_\beta}

Like in the bosonic case, we can define various classes of 
correlation functions. The ``physical'' correlators are now set up as
\ba
 \Pi^{>}_{\alpha\beta}(Q) & \equiv & 
 \I \Bigl\langle \A(x) \B(0) \Bigl\rangle
 \;,   
 \la{fL}
 \\
 \Pi^{<}_{\alpha\beta}(Q) & \equiv & 
 \I \Bigl\langle - \B(0)  \A(x) \Bigl\rangle
 \;,   
 \la{fS}
 \\
 \rho_{\alpha\beta}(Q) & \equiv & 
 \I \Bigl\langle \fr12 \Bigl\{ \A(x) , \B(0) \Bigr\} \Bigl\rangle
 \;,   
 \la{frho}
\ea
where $\rho_{\alpha\beta}$ is the spectral function, 
while retarded and advanced correlators can be defined as 
\ba
 \Pi^R_{\alpha\beta}(Q) & \equiv & 
 i \I \Bigl\langle \Bigl\{ \A(x) , \B(0) \Bigr\} \theta(t) \Bigl\rangle
 \;, 
 \la{fR}
 \\
 \Pi^A_{\alpha\beta}(Q) & \equiv & 
 i \I \Bigl\langle  - \Bigl\{ \A(x) , \B(0) \Bigr\} \theta(-t) \Bigl\rangle
 \;. 
 \la{fA} 
\ea
On the other hand, the time-ordered correlation function reads 
\ba
 \Pi^T_{\alpha\beta}(Q) & \equiv & 
  \I \Bigl\langle \A(x)  \B(0) \theta(t)
                 -\B(0)  \A(x) \theta(-t) \Bigl\rangle
 \;, \la{fT}
\ea
while the Euclidean correlator is
\ba
 \Pi^E_{\alpha\beta}(\tilde Q) & \equiv & 
 \int_0^\beta\!{\rm d}\tau \int \! {\rm d}^3 \vec{x}\,
  e^{i \tilde Q \cdot \tilde x}
  \Bigl\langle \A(\tilde x)  \B(0) \Bigl\rangle
 \;.
 \la{fE}
\ea
Note again that the Euclidean correlator is time-ordered by definition, 
and can be computed with standard imaginary-time functional integrals 
in the Matsubara formalism. 

Like in the bosonic case, all of the correlation functions defined 
can be expressed in terms of the spectral
function, which in turn can be determined as a certain analytic 
continuation of the Euclidean correlator. First, 
inserting sets of energy eigenstates, we obtain  
the KMS-relation in Fourier-space,  
$
 \Pi^{<}_{\alpha\beta}(Q) 
 = - e^{-\beta q^0} \Pi^{>}_{\alpha\beta}(Q)
$.
Then 
$
 \rho_{\alpha\beta}(Q)
 = [\Pi^{>}_{\alpha\beta}(Q) - \Pi^{<}_{\alpha\beta}(Q)]/2
$ and, conversely, 
\be
 \Pi^{>}_{\alpha\beta}(Q) = 2[1 - \nF{}(q^0)] \rho_{\alpha\beta}(Q)
 \;, \quad
 \Pi^{<}_{\alpha\beta}(Q) = - 2 \nF{}(q^0) \rho_{\alpha\beta}(Q)
 \;, \la{fLSrel}
\ee
where $\nF{}(x)\equiv 1/[\exp(\beta x) + 1]$. Inserting the representation
of \eq\nr{theta} into the definitions of $\Pi^R$, $\Pi^A$ produces
\be
 \Pi^R_{\alpha\beta}(Q) 
 =  \int_{-\infty}^{\infty} \! \frac{{\rm d}\omega}{\pi} 
 \frac{\rho_{\alpha\beta}(\omega,\vec{q})}{\omega -q^0- i 0^+}
 \;, \quad 
 \Pi^A_{\alpha\beta}(Q) 
 =  \int_{-\infty}^{\infty} \! \frac{{\rm d}\omega}{\pi} 
 \frac{\rho_{\alpha\beta}(\omega,\vec{q})}{\omega -q^0+ i 0^+}
 \;.  \la{fRrhorel}
\ee
Proceeding similarly with $\Pi^T$ and making use of \eq\nr{delta},  
we obtain
\be
 \Pi^T_{\alpha\beta}(Q) = 
 \int_{-\infty}^{\infty} \! \frac{{\rm d}\omega}{\pi} 
 \frac{i\rho_{\alpha\beta}(\omega,\vec{q})}{q^0 - \omega + i 0^+} - 
 2 \rho_{\alpha\beta}(q^0,\vec{q}) \nF{}(q^0)
 \;. 
\ee
Finally, writing the argument inside the $\tau$-integration
in \eq\nr{fE} as a Wick rotation of
the inverse Fourier transform of the left-hand side of \eq\nr{fL}, 
inserting \eq\nr{fLSrel}, and changing orders of integration, we get 
\ba
 \Pi^E_{\alpha\beta}(\tilde Q) & = & 
 \int_0^\beta \! {\rm d}\tau\, e^{i \qq \tau}
 \int_{-\infty}^{\infty} \frac{{\rm d}\omega }{2\pi} e^{-\omega \tau} 
 \Pi^{>}_{\alpha\beta}(\omega,\vec{q}) =
 \int_{-\infty}^{\infty} \! \frac{{\rm d}\omega }{\pi} 
 \frac{\rho_{\alpha\beta}(\omega,\vec{q})}{\omega - i \qq}
 \;. \la{fErhorel}
\ea
Like in the bosonic case, this relation can be 
inverted by making use of \eq\nr{delta}, 
\be
 \rho(q^0,\vec{q}) = \frac{1}{2i} \disc \Pi^E(\qq\to -i q^0,\vec{q})
 \;. \la{ffinal}
\ee

We also recall that fermionic Matsubara sums can be carried out through
\ba
  T \sum_{\omega_\rmi{f}} 
 \frac{i \omega_\rmi{f} c + d }{\omega_\rmi{f}^2 + E^2}  
 e^{i \omega_\rmi{f} \tau} 
  & \equiv & 
  (c \partial_\tau + d ) T \sum_{\omega_\rmi{f}} 
 \frac{ e^{i \omega_\rmi{f} \tau} }{\omega_\rmi{f}^2 + E^2}  
 \\ 
 & = & 
 \frac{\nF{}(E)}{2 E} \Bigl[ 
  (- cE + d) e^{(\beta - \tau) E} -(c E + d ) e ^{\tau E}
 \Bigr] \;, 
 \la{fsum}
\ea
where $\omega_\rmi{f} = 2 \pi T (n + \fr12)$,
with $n$ an integer, and we assumed $0 < \tau < \beta$;
and that a typical integration yields
\be
 \int_0^\beta\! {\rm d}\tau \,
 e^{-\tau (i \omega_\rmi{f} + \Delta)}
 = 
 \frac{1 + e^{-\beta\Delta}}{i \omega_\rmi{f} + \Delta} 
 \;. \la{fint}
\ee

%
\section{An alternative derivation of \eq\nr{master}}

We present in this Appendix an alternative derivation
(following Ref.~\cite{mlb}, for example) of \eq\nr{master}, 
which is technically somewhat simpler
than the one in the main text, but with the price of containing
a few heuristic steps. The end result is nevertheless identical. 

The starting point is the interaction Hamiltonian in the phase
with broken electroweak symmetry, \eq\nr{HI}.
Consider now an initial state 
$
 | \mathcal{I} \rangle = | \rm{i} \rangle \otimes | 0 \rangle
$
and a final state 
$
 | \mathcal{F} \rangle = $ $| \rm{f} \rangle \otimes | I;\vec{q},s \rangle
$, 
where the right subspace contains the sterile neutrinos, and 
\be
 | I;\vec{q},s \rangle \equiv \hat a^{\dagger}_{I;\vec{q},s} | 0 \rangle
 \;, \quad
 \langle I;\vec{q},s | = \langle 0 | \hat a_{I;\vec{q},s}
 \;.
\ee
The transition matrix element can immediately be written down, 
\ba
 T_{\mathcal{FI}} & = & \langle \mathcal{F} | 
 \int \! {\rm d} t\, \hat H_I(t) | \mathcal{I} \rangle
 = 
 \int\! {\rm d}t\, {\rm d}^3\vec{x} \,
 \frac{e^{i Q\cdot x}}{\sqrt{\raise-0.2ex\hbox{$(2\pi)^3 2 q^0$}}}
 \langle {\rm f} | \hat J_{I;\vec{q},s}(x) | {\rm i} \rangle
 \;, 
\ea
where we inserted \eq\nr{HI2}, 
and $\hat J_{I;\vec{q},s}$ is from \eq\nr{J}. The production rate
can then be obtained by summing over all initial states, with their
proper Boltzmann weights, and over all allowed final states: 
\ba
 \frac{{\rm d} N_I(x,\vec{q})}{{\rm d}^4 x\, {\rm d}^3 \vec{q}} & = & 
 \lim_{V,\Delta t\to\infty} \frac{1}{V \Delta t} 
 \sum_{s=\pm 1} \sum_{\rm f,i} \frac{e^{-\beta E_i}}{Z}
 |T_{\mathcal{FI}}|^2
 \\ 
 & = &  
 \frac{1}{(2\pi)^3 2 q^0}
 \int\! {\rm d}t\, {\rm d}^3\vec{x} \,
 e^{i Q \cdot x} 
\sum_{s=\pm 1} 
 \Bigl \langle \hat{J}^\dagger_{I;\vec{q},s}(x)
 \hat{J}^{\mbox{ }}_{I;\vec{q},s}(0) \Bigr\rangle
 \;,
\ea
where $V$ is the volume, $\Delta t$ is the time interval, 
$Z$ the partition function, we defined a thermal average by
$
 \langle ... \rangle \equiv Z^{-1} \tr[\exp(-\beta \hat H_\rmi{MSM})(...)]
$,  
and made use of translational invariance. 

It remains to repeat steps (ii), (iii) in the paragraph 
following \eq\nr{J}. We thus arrive directly at \eq\nr{raw}.



\begin{thebibliography}{99}


\bibitem{wdm}
  S.~Dodelson and L.M.~Widrow,
  Phys.\ Rev.\ Lett.\  {72} (1994) 17
  [hep-ph/9303287].

\bibitem{ot}
  K.A.~Olive and M.S.~Turner,
  Phys.\ Rev.\ D {25} (1982) 213.
  
  \bibitem{astro}
  A.~Kusenko and G.~Segr\`e,
  Phys.\ Rev.\ Lett.\  {77} (1996) 4872
  [hep-ph/9606428];
%
  G.M.~Fuller, A.~Kusenko, I.~Mocioiu and S.~Pascoli,
  Phys.\ Rev.\ D {68} (2003) 103002
  [astro-ph/0307267];
%
  M.~Barkovich, J.C.~D'Olivo and R.~Montemayor,
  Phys.\ Rev.\ D {70} (2004) 043005
  [hep-ph/0402259];
%
  P.L.~Biermann and A.~Kusenko,
  Phys.\ Rev.\ Lett.\  {96} (2006) 091301
  [astro-ph/0601004].

\bibitem{numsm}
  T.~Asaka, S.~Blanchet and M.~Shaposhnikov,
  Phys.\ Lett.\ B {631} (2005) 151
  [hep-ph/0503065].

\bibitem{osc}
  E.K.~Akhmedov, V.A.~Rubakov and A.Y.~Smirnov,
  Phys.\ Rev.\ Lett.\  {81} (1998) 1359
  [hep-ph/9803255].

\bibitem{baryo}
  T.~Asaka and M.~Shaposhnikov,
  Phys.\ Lett.\ B {620} (2005) 17
  [hep-ph/0505013].
  
\bibitem{st}
  M.~Shaposhnikov and I.~Tkachev,
  hep-ph/0604236.
  
  
\bibitem{lss}
  S.H.~Hansen, J.~Lesgourgues, S.~Pastor and J.~Silk,
  Mon.\ Not.\ Roy.\ Astron.\ Soc.\  {333} (2002) 544
  [astro-ph/0106108];
%
  M.~Viel, J.~Lesgourgues, M.G.~Haehnelt, S.~Matarrese and A.~Riotto,
  Phys.\ Rev.\ D {71} (2005) 063534
  [astro-ph/0501562];
%
  K.~Abazajian,
  Phys.\ Rev.\ D {73} (2006) 063513
  [astro-ph/0512631].

\bibitem{seljak}  
  U.~Seljak, A.~Makarov, P.~McDonald and H.~Trac,
  astro-ph/0602430.
  
\bibitem{life} 
  A.D.~Dolgov and S.H.~Hansen,
  Astropart.\ Phys.\  {16} (2002) 339 
  [hep-ph/0009083].
  
\bibitem{Abazajian:2001vt}
  K.~Abazajian, G.M.~Fuller and W.H.~Tucker,
  Astrophys.\ J.\  {562} (2001) 593
  [astro-ph/0106002].
  
\bibitem{xray}
  A.~Boyarsky, A.~Neronov, O.~Ruchayskiy and M.~Shaposhnikov,
  astro-ph/0512509;
%
  astro-ph/0603368.

\bibitem{boyagal}  
  A.~Boyarsky, A.~Neronov, O.~Ruchayskiy, M.~Shaposhnikov and I.~Tkachev,
  astro-ph/0603660.

\bibitem{hansen}  
  S.~Riemer-S{\o}rensen, S.H.~Hansen and K.~Pedersen,
  astro-ph/0603661.
%

\bibitem{ReSigmamu}
  X.~Shi and G.M.~Fuller,
  Phys.\ Rev.\ Lett.\  {82} (1999) 2832
  [astro-ph/9810076].


\bibitem{late}
  T.~Asaka, A.~Kusenko and  M.~Shaposhnikov,
  hep-ph/0602150.

\bibitem{oldwdm}
  K.~Abazajian, G.M.~Fuller and M.~Patel,
  Phys.\ Rev.\ D {64} (2001) 023501 
  [astro-ph/0101524];
%
  K.N.~Abazajian and G.M.~Fuller,
  Phys.\ Rev.\ D {66} (2002) 023526
  [astro-ph/0204293].
  
\bibitem{oldwdm1}  K.~Abazajian,
  Phys.\ Rev.\ D {73} (2006) 063506 
  [astro-ph/0511630].
  
\bibitem{ak}
  K.~Abazajian and S.M.~Koushiappas,
  astro-ph/0605271.
  
  
\bibitem{bd}
  R.~Barbieri and A.~Dolgov,
  Phys.\ Lett.\ B {237} (1990) 440;
%
  Nucl.\ Phys.\ B {349} (1991) 743.
  
\bibitem{phen}
  M.~Laine and Y.~Schr\"oder,
  Phys.\ Rev.\ D {73} (2006) 085009
  [hep-ph/0603048].

  
\bibitem{dilepton}
  L.D.~McLerran and T.~Toimela,
  Phys.\ Rev.\ D {31} (1985) 545;
%
  H.A.~Weldon,
  Phys.\ Rev.\ D {42} (1990) 2384;
%
  C.~Gale and J.I.~Kapusta,
  Nucl.\ Phys.\ B {357} (1991) 65.

\bibitem{nunu}
  V.N.~Tsytovich, 
  Sov.\ Phys.\ JETP 13 (1961) 1249 
  [Zh.\ Eksp.\ Teor.\ Fiz.\ 40 (1961) 1775]; 
%
  J.B.~Adams, M.A.~Ruderman and C.-H.~Woo, 
  Phys.\ Rev.\ 129 (1963) 1383; 
%
  E.~Braaten and D.~Segel,
  Phys.\ Rev.\ D {48} (1993) 1478
  [hep-ph/9302213].
  
\bibitem{old}
  A.L.~Fetter and J.D.~Walecka, 
  {\it Quantum Theory of Many-Particle Systems}
  (McGraw-Hill, New York, 1971); 
  %
  S.~Doniach and E.H.~Sondheimer, 
  {\it Green's Functions for Solid State Physicists}
  (Benjamin, Reading, 1974);
  %
  J.W.~Negele and H.~Orland,
  {\it Quantum Many Particle Systems}
  (Addison-Wesley, Redwood City, 1988).
  
\bibitem{mlb}
  M. Le Bellac, {\it Thermal Field Theory}
  (Cambridge University Press, Cambridge, 2000).

\bibitem{expansion}
  J. Bernstein, {\it Kinetic Theory in the Expanding Universe}
  (Cambridge University Press, Cambridge, 1988);
%
  E.W.~Kolb and M.S.~Turner, {\it The Early Universe}
  (Addison-Wesley, Reading, 1990).

\bibitem{weldon}
  H.A.~Weldon,
  Phys.\ Rev.\ D {26} (1982) 2789.



\bibitem{ReSigmaold}
  D.~N\"otzold and G.~Raffelt,
  Nucl.\ Phys.\ B {307} (1988) 924;
%
  K.~Enqvist, K.~Kainulainen and J.~Maalampi,
  Nucl.\ Phys.\ B {349} (1991) 754;
%
  J.C.~D'Olivo, J.F.~Nieves and M.~Torres,
  Phys.\ Rev.\ D {46} (1992) 1172.

\bibitem{ReSigma}
  C.~Quimbay and S.~Vargas-Castrill\'on,
  Nucl.\ Phys.\ B {451} (1995) 265
  [hep-ph/9504410].

\bibitem{nrqcd3}
  M.~Laine and M.~Veps\"al\"ainen,
  JHEP {02} (2004) 004
  [hep-ph/0311268].

\bibitem{sums}
  A.I.~Bochkarev and M.E.~Shaposhnikov,
  Nucl.\ Phys.\ B {268} (1986) 220;
%
  A.~I.~Bochkarev,
  Sov.\ J.\ Nucl.\ Phys.\  {45} (1987) 517
  [Yad.\ Fiz.\  {45} (1987) 832];
%
  C.A.~Dominguez and M.~Loewe,
  Phys.\ Lett.\ B {233} (1989) 201;
%
  R.J.~Furnstahl, T.~Hatsuda and S.H.~Lee,
  Phys.\ Rev.\ D {42} (1990) 1744;
%
  T.~Hatsuda, Y.~Koike and S.H.~Lee,
  Nucl.\ Phys.\ B {394} (1993) 221.

\bibitem{bp}
  E.~Braaten, R.D.~Pisarski and T.C.~Yuan,
  Phys.\ Rev.\ Lett.\  {64} (1990) 2242.

\bibitem{htl}
  P.~Aurenche, F.~Gelis, G.D.~Moore and H.~Zaraket,
  JHEP {12} (2002) 006
  [hep-ph/0211036];
%
  P.~Arnold, G.D.~Moore and L.G.~Yaffe,
  JHEP {12} (2001) 009
  [hep-ph/0111107];
%
  JHEP {01} (2003) 030
  [hep-ph/0209353];
%
  JHEP {05} (2003) 051
  [hep-ph/0302165];
%
and references therein.


\bibitem{acont}
  G.~Cuniberti, E.~De Micheli and G.A.~Viano,
  Commun.\ Math.\ Phys.\  {216} (2001) 59.

\bibitem{mem}
  M.~Asakawa, T.~Hatsuda and Y.~Nakahara,
  Prog.\ Part.\ Nucl.\ Phys.\  {46} (2001) 459
  [hep-lat/0011040];
%
  F.~Karsch, E.~Laermann, P.~Petreczky, S.~Stickan and I.~Wetzorke,
  Phys.\ Lett.\ B {530} (2002) 147
  [hep-lat/0110208].

\bibitem{mem2}
  G.~Aarts and J.M.~Mart\'{\i}nez Resco,
  JHEP {04} (2002) 053
  [hep-ph/0203177].

\bibitem{xpt}
  J.~Gasser and H.~Leutwyler,
  Phys.\ Lett.\ B {184} (1987) 83.

\bibitem{ubw}
  M.~Urban, M.~Buballa and J.~Wambach,
  Phys.\ Rev.\ Lett.\  {88} (2002) 042002
  [nucl-th/0110005];
%
  D.~Fern\'andez-Fraile and A.~G\'omez Nicola,
  Phys.\ Rev.\ D {73} (2006) 045025
  [hep-ph/0512283];
%
  and references therein.

\bibitem{hisuscs}
  M.~Asakawa, U.W.~Heinz and B.~M\"uller,
  Phys.\ Rev.\ Lett.\  {85} (2000) 2072
  [hep-ph/0003169];
 %
  S.~Jeon and V.~Koch,
  Phys.\ Rev.\ Lett.\  {85} (2000) 2076
  [hep-ph/0003168].

\bibitem{av}
  A.~Vuorinen,
  Phys.\ Rev.\ D {67} (2003) 074032
  [hep-ph/0212283].

\bibitem{bir}
  J.P.~Blaizot, E.~Iancu and A.~Rebhan,
  Phys.\ Lett.\ B {523} (2001) 143
  [hep-ph/0110369].

\bibitem{suscs}
  R.V.~Gavai, S.~Gupta and P.~Majumdar,
  Phys.\ Rev.\ D {65} (2002) 054506
  [hep-lat/0110032];
%
  C.~Bernard {\it et al.}  [MILC Collaboration],
  Phys.\ Rev.\ D {71} (2005) 034504
  [hep-lat/0405029];
%
  C.R.~Allton {\it et al.},
  Phys.\ Rev.\ D {71} (2005) 054508
  [hep-lat/0501030];
%
  R.V.~Gavai and S.~Gupta,
  Phys.\ Rev.\ D {73} (2006) 014004
  [hep-lat/0510044].

\bibitem{ay}
  P.~Arnold and L.G.~Yaffe,
  Phys.\ Rev.\ D {57} (1998) 1178
  [hep-ph/9709449].

\bibitem{weldon2}
  H.A.~Weldon,
  Phys.\ Rev.\ D {28} (1983) 2007.

\bibitem{damping}
  K.~Enqvist, K.~Kainulainen and M.J.~Thomson,
  Nucl.\ Phys.\ B {373} (1992) 498;
%
  P.~Langacker and J.~Liu,
  Phys.\ Rev.\ D {46} (1992) 4140
  [hep-ph/9206209];
%
  E.S.~Tututi, M.~Torres and J.C.~D'Olivo,
  Phys.\ Rev.\ D {66} (2002) 043001
  [hep-th/0209006].


\end{thebibliography}
\end{document}